\documentclass[]{aa}
\usepackage{soul}
\usepackage{natbib}
\def\kms{\hbox{km$\;$s$^{-1}$}}
\def\halpha{H$\mathrm{\alpha}$}
\def\cak{\hbox{\ion{Ca}{ii}~K}}
\def\Mgk{\hbox{\ion{Mg}{ii}~k}}
\usepackage{graphicx}
\usepackage{epstopdf}
\usepackage{lipsum}
\usepackage{mwe}
\usepackage{caption}
\usepackage{subcaption}
\usepackage{txfonts}
\usepackage[bookmarks = true,colorlinks = true,linkcolor = red,urlcolor = blue,citecolor = blue,pdftex,bookmarks = true,linktocpage = true,hyperindex = true,breaklinks]{hyperref}
\begin{document} 

  \title{Spicules and downflows in the solar chromosphere}
  \author{Souvik Bose \inst{1}$^,$\inst{2}
  \and 
  Jayant Joshi\inst{1}$^,$\inst{2}
  \and
  Vasco M.J. Henriques\inst{1}$^,$\inst{2}
  \and
  Luc Rouppe van der Voort\inst{1}$^,$\inst{2}
          }

  \institute{Institute of Theoretical Astrophysics, University of Oslo, P.O. Box 1029 Blindern, NO-0315 Oslo, Norway
          \and
    Rosseland Centre for Solar Physics, University of Oslo, P.O. Box 1029 Blindern, NO-0315 Oslo, Norway\\
             \email{souvik.bose@astro.uio.no}
             }

  \date{XXXX; accepted XXXX}

  \abstract
   {High speed downflows have been observed in the solar transition region (TR) and lower corona for many decades. Despite their abundance, it has been hard to find signatures of such downflows in the solar chromosphere. }
   {In this work, we target an enhanced network region which shows ample occurrences of rapid spicular downflows in the \halpha\ spectral line that could potentially be linked to high-speed TR downflowing counterparts.}
   {We used the $k$-means algorithm to classify the spectral profiles of on-disk spicules in \halpha{} and \cak{} data observed from the Swedish 1-m Solar Telescope (SST) and employed an automated detection method based on advanced morphological image processing operations to detect such downflowing features, in conjunction with rapid blue shifted and red shifted excursions (RBEs and RREs).}
   {We report the existence of a new category of RREs (termed as downflowing RRE) for the first time that, contrary to earlier interpretation, are associated with chromospheric field aligned downflows moving towards the strong magnetic field regions. Statistical analysis performed on nearly 20,000 RBEs and 15,000 RREs (including the downflowing counterparts), detected in our 97~min long dataset, shows that the downflowing RREs are very similar to RBEs and RREs except for their oppositely directed plane-of-sky motion. Furthermore, we also find that RBEs, RREs and downflowing RREs can be represented by a wide range of spectral profiles with varying Doppler offsets, and \halpha{} line core widths, both along and perpendicular to the spicule axis, that causes them to be associated with multiple substructures that evolve together.}
   {We speculate that these rapid plasma downflows could well be the chromospheric counterparts of the commonly observed TR downflows. }

  \authorrunning{Bose et al.}

\maketitle


\section{Introduction}
\label{Section:intro}

\begin{figure*}
   \centering
   \includegraphics[width=1.05\textwidth]{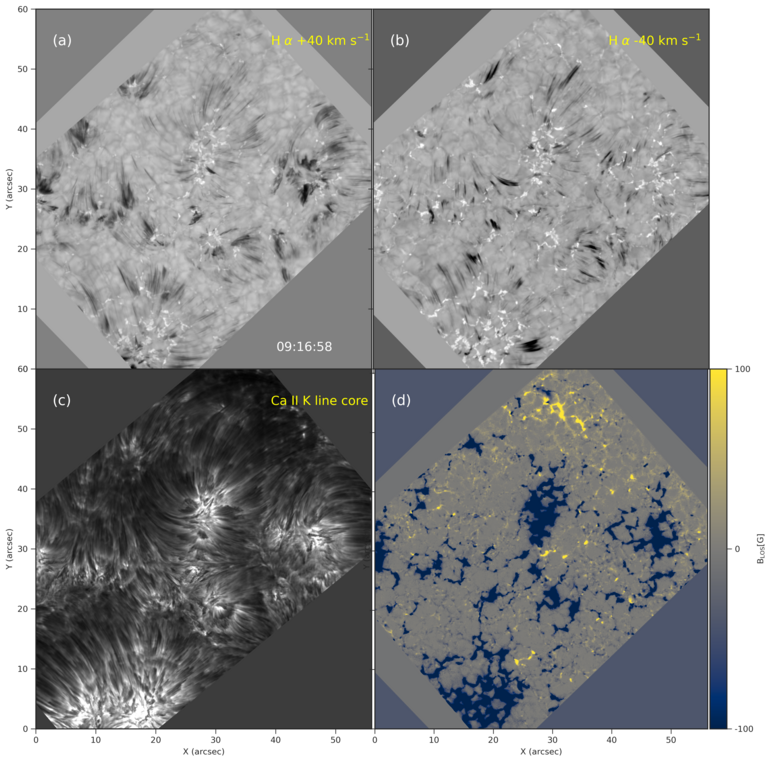}
   \caption{Full FOV of the SST dataset observed on 25 May 2017 at 09:16:58 UTC. Panels~(a) and (b) show the \halpha{} red wing and blue wing images observed with CRISP at a Doppler offset of +40~\kms and $-$40~\kms{}, respectively. Spicules are seen in the two panels as dark and elongated thread-like structures. Panel~(c) shows a dense chromospheric canopy of fibrils in the corresponding CHROMIS \cak{} line core image and (d) shows the map of the LOS magnetic field (B$_{\mathrm{LOS}}$) saturated between $\pm$100~G derived from inversions of the \ion{Fe}{I}~6301 and 6302~\AA\ spectral profiles. The direction to solar North is pointing upwards. An animation of this figure is available at \url{https://www.dropbox.com/s/nxnz5twzczjwaac/Context_movie1.mp4?dl=0}
   } 
        \label{figure:Context_v1}%
    \end{figure*}

Downflows are known to commonly occur in the solar atmosphere. High speed downflows have been observed in the transition region (TR) that can sometimes last from several hours to even several days \citep[e.g.][]{1981ApJ...251L.115G,1982SoPh...77...77D}. Moreover, stronger plasma downflows with speeds ranging from $60$--$200$~\kms{} have been observed over or in the close vicinity of active regions and quiet Sun alike, with the latter revealing relatively weaker downflows \citep{1976ApJ...205L.177D,1988ApJ...334.1066K,1997SoPh..175..349B,1999ApJ...522.1148P}. Decades of observations have revealed the prevalence of predominant downflows (or red shifts) in the spectral lines of the TR \citep{1976ApJ...205L.177D,1981ApJ...249..720S,1982SoPh...77...77D,1999ApJ...522.1148P,2011A&A...534A..90D}. These downflows are considered to play an important role in the mass and energy balance of the solar atmosphere and are crucial to further understand the physics of the lower solar corona.

These downflows are predominantly seen mainly in the lower coronal and TR passbands. \citet{1997SoPh..175..349B,1999ApJ...522.1148P} reported that most of the high speed quiet Sun TR downflows usually vanish at chromospheric temperatures. \cite{2016A&A...587A..20C} have found supersonic downflows in the TR of sunspots, which are highly intermittent in time and location \citep{2020A&A...636A..35N}.  Statistical analysis of such downflows in $48$ sunspots by \citet{2018ApJ...859..158S} show that at most half of them show signatures in the upper chromospheric spectral lines such as \Mgk{}, whereas the rest are limited to the TR. Sunspots (and active regions in general) occupy only a very small fraction of the solar surface compared to the non active regions. Therefore, it has remained a mystery as to what happens to the remaining high speed downflows in the chromosphere that are so predominant in the TR of non active regions. Why have their signatures been so elusive in the chromosphere? A possible reason for such lack of evidence could be because the plasma density in the chromosphere is several orders of magnitude higher compared to the lower corona or the TR that causes these downflows to simply breakdown as soon as they reach chromospheric heights. 

The chromosphere is a layer of the Sun's atmosphere that is sandwiched between the visible photosphere and the million degree corona. It is dominated by a myriad of different features such as spicules, dynamic fibrils, and filaments to name a few. All these features contribute towards the vigorous dynamical changes that the chromosphere experiences on time scales ranging from seconds to minutes. The chromosphere also plays a crucial role in mass loading and heating the solar corona since all the non-thermal energy that is responsible for these mechanisms, propagates through the chromosphere. Only a small fraction of this energy escapes into the corona or the solar wind while the majority remains trapped within \citep{2019ARA&A..57..189C}.

One of the most abundant and ubiquitous features that appear in the chromosphere are spicules. Spicules are thin, elongated, thread-like, and highly dynamic structures that are omnipresent in the solar atmosphere both in active and non active regions.
To our knowledge, the first lithographed drawings resembling spicules were performed by Pietro Tachinni 
\citep[see, for historic references and a modern reproduction, ][]{2006MSAIS...9...28C}
and Angelo Secchi \citep{1871AdPd}, both in April 1871, with the drawings in the second edition of Le Soleil \citep{1877arnp.book.....S} being better known examples of these early observations. Spicules are visible all over the solar surface when observed in chromospheric spectral lines and in some cases are known to reach coronal temperatures. They have been of great interest to the solar physics community for a long time. An overview of some of such work can be found at \cite{1968SoPh....3..367B,2000SoPh..196...79S,Tsiropoula_2012,2019PASJ...71R...1H}.


The discovery of a more elusive but energetic category of spicules by \citet{Bart_2007_PASJ} divided them into two different classes: type-I and type-II, and stirred a fierce debate in the community regarding their existence, formation mechanisms and role in coronal heating. The type-Is are mainly used to refer to the dynamic fibrils found in active regions or mottles in the quiet Sun, and they usually appear in the close vicinity of strong magnetic fields \citep{2004Natur.430..536D}. Further, they display characteristic parabolic paths in \halpha{} line core space-time diagrams with typical up and down motions of the order of  $10$--$40$~\kms{}, lifetimes between $3$--$5$~min and quasi-periodicities of roughly the same time periods \citep{Luc_2007}. Advanced numerical simulations by \citet{2006ApJ...647L..73H,2007ApJ...666.1277H} show remarkable similarities with the observations described and they firmly established that type-I spicules are formed due to the leakage of photospheric magnetoacoustic oscillations into the solar chromosphere. The type-II spicules, on the other hand, are comparatively more dynamic with vigorous sideways motions, high apparent speeds (80--300 \kms{}) and shorter lifetimes (\textasciitilde $1$--$3$~minutes) \citep[][]{Bart_2007_PASJ,Tiago_2012,2016ApJ...824...65P}. Moreover, they are known to be heated as they propagate beyond the chromosphere and become visible in TR \citep{Tiago_2014_heat,Luc_2015}
and coronal passbands in both active regions and the quiet Sun  \citep{2011Sci...331...55D,Vasco_2016,2016ApJ...830..133K,2019Sci...366..890S}. These qualities render the importance of attributing type-II spicules towards mass loading and heating the solar corona
\citep[see, e.g.,][]{Juan_2017_Science,2018SoPh..293...56K,Juan_2018}, however, their detailed physical processes remain far from known.

The spectral signatures of the on-disk counterparts of type-II spicules were observed for the first time by \citet{2008ApJ...679L.167L}. They reported sudden rapid excursions in the blue wing of \ion{Ca}{II}~8542~\AA\ spectral line that led them to be termed as rapid blue shifted excursions (RBEs). Later, \citet{Luc_2009,Sekse_2012,2016ApJ...824...65P,my_paper_3} observed high resolution on-disk images and spectra in the chromospheric \halpha{}, \ion{Ca}{ii}~8542~\AA, and \cak{} lines associated with them and established beyond doubt that RBEs are indeed the on-disk counterparts of the earlier known type-II spicules. 

\citet{Bart_3_motions} reported the presence of torsional motions in type-II spicules that explained the occurrence of spicular bushes in the red wing images of chromospheric spectral lines, that were morphologically similar to their blue counterparts. Later, \citet{2013ApJ...769...44S} also described the existence of red wing counterparts of RBEs in \halpha{} and \ion{Ca}{ii}~8542\AA\ with very similar properties as the former and termed them rapid red shifted excursions (RREs). Next, \citet{2015ApJ...802...26K,Luc_2015} and more recently \citet{my_paper_3} also confirmed their existence, both in the chromosphere and the TR, with high resolution on-disk observations. 

The complex twisting and swaying found in type-II spicules, in addition to the flows along the chromospheric magnetic field lines, are important characteristics of spicules that represents outward propagating Alfv{\'e}nic waves which can be of the order of several hundred \kms\, and can cause heating in the hotter TR lines as they propagate \citep{2014Sci...346D.315D}. Like \citet{Bart_3_motions}, \citet{2013ApJ...769...44S} and later \citet{2015ApJ...802...26K} used the transverse motion of spicules to argue for the existence and appearance of RREs. They interpreted that RREs (though less abundant), like RBEs, are a manifestation of the same phenomenon with very similar statistical properties and occur when the latter harbor such complex motions that can sometimes result in a net red shift when observed on the disk. Therefore, RBEs can transition to RREs and vice-versa depending on the orientation between the line-of-sight (LOS) and transverse motion of the structures. Moreover, the torsional motions can also cause RBEs and RREs to be in close association with each other \citep{2013ApJ...769...44S,2014Sci...346D.315D}. Both RBEs and RREs are generally seen to originate in the vicinity of strong network regions with enhanced magnetic fields, and appear to propagate away as they evolve \citep{Luc_2009,2009ApJ...707..524M,2013ApJ...769...44S,2014Sci...346D.315D,2015ApJ...802...26K}

Movies of high-resolution \halpha\ blue wing filtergrams at Doppler offsets $\gtrsim$30~\kms\ are generally dominated by RBEs that appear to move away from the magnetic network. Corresponding red wing movies at equivalent Doppler offset, are not nearly as much dominated by outward moving RREs, but also show elongated absorption features that appear to be moving downward to the magnetic network. 
In this paper, with the help of high resolution on-disk observations from the Swedish 1-m Solar Telescope \citep[SST,][]{2003SPIE.4853..341S}, we aim to characterize these returning flows that appear morphologically similar to RBEs and RREs, but seem to be downflowing in nature.
%
In the following sections, we describe the methods undertaken to detect and further investigate their dynamical characteristics and spatio-temporal evolution. We also discuss their interpretation and the possible physical processes that could be responsible for their appearance and suggest that these downflows could be a representative of the chromospheric counterparts of the lower coronal and TR downflows.


\section{Observations and data reduction}
\label{Section:OBS}

We observed an enhanced network region close to disk center in a coordinated SST and IRIS campaign on 25 May 2017 shown in Fig.~\ref{figure:Context_v1}. The heliocentric coordinates were  $(x,y)=(45\arcsec,-93\arcsec)$ with corresponding observing angle $\mu=\cos \theta=0.99$ ($\theta$ being the heliocentric angle). The temporal duration was close to 97 min starting from 09:12 UT until 10:49 UT. We acquired imaging spectroscopic data in \halpha{} and spectropolarimetric data in \ion{Fe}{I}~6302~\AA\ from the CRisp Imaging SpectroPolarimeter \citep[CRISP,][]{Crisp_2008}. The CHROmospheric Imaging Spectrometer (CHROMIS) was used to obtain imaging spectroscopic data in \cak{}. Both CRISP and CHROMIS are tunable Fabry-P{\'e}rot instruments installed at the SST with the first light of the latter being in 2016. CRISP sampled \halpha{} at 32 wavelength positions between $\pm$1.85~\AA\ and the  \ion{Fe}{I}~6301 and 6302~\AA\ line pair at 16 wavelength positions respectively, with a temporal cadence of 19.6~s and a spatial sampling of 0\farcs058. The \ion{Fe}{I} spectral profiles were subjected to an early version of a robust Milne-Eddington (ME) inversion scheme based on a parallel \verb|C++|/\verb|Python| implementation\footnote{ https://github.com/jaimedelacruz/pyMilne} \citep{2019A&A...631A.153D}. We use the LOS magnetic field component derived from these inversions in our analysis as shown in panel~(d) of Fig.~\ref{figure:Context_v1}. CHROMIS sampled \cak{} at 41 wavelength positions within $\pm$1.28~\AA\ with 63.5~m\AA\ steps. Furthermore, a continuum position was sampled at 4000~\AA. The CRISP \halpha{} red-wing and blue-wing images for the full field-of-view (FOV) at +40~\kms{} and $-$40~\kms{} are respectively shown in panels~(a) and (b) of Fig.~\ref{figure:Context_v1}. The corresponding CHROMIS \cak{} line core image is shown in panel~(c). The temporal cadence and the spatial scale of this data were 13.6~s and 0\farcs038. High spatial resolution down to the telescopic diffraction limit (given by $\lambda/\mathrm{D}$ = 0\farcs08 at 3934\AA) was achieved through excellent seeing conditions, the SST adaptive optics system \citep{2019A&A...626A..55S}, 
and the Multi-Object Multi-Frame Blind Deconvolution \citep[MOMFBD,][]{vannoort2005MOMFBD} image restoration technique.

We used the CRISPRED data reduction pipeline \citep{2015A&A...573A..40D} and an early version of the CHROMIS pipeline \citep{2018arXiv180403030L} for further data reduction while including the spectral consistency method described in \cite{2012A&A...548A.114H}. The images from both the instruments were de-rotated to account for diurnal field rotation, aligned and de-stretched to remove warping due to seeing effects before making it ready for scientific analyses. Later, both the CRISP and the CHROMIS datasets were co-aligned by cross-correlating the corresponding photospheric wideband channels and were rotated to align the direction to solar North along the $y$-axis. 
The wideband channel for CRISP has a full-width at half maximum (FWHM)=4.9~\AA\ centered at \halpha, whereas for CHROMIS the FWHM equals 13.2~\AA\ centered at 3950~\AA\ between the \ion{Ca}{II}~H and \ion{Ca}{II}~K lines. 

These observations were first presented by \citet{my_paper_3} where the cotemporal IRIS observations were included in the analysis. The orientation and total area covered by the dataset matches the IRIS observations. This data will be made available in the future as a part of the SST and IRIS database \citep{2020A&A...641A.146R}.
A brief overview of the dataset and targetted features of interest are given in the following section.

\subsection{Overview of the dataset}

Figure~\ref{figure:Context_v1} shows an overview of the FOV and the qualification, enhanced network, is illustrated by extended patches of strong magnetic field that are predominately of negative magnetic polarity in the $B_\mathrm{LOS}$ map.
The \cak{} line core image is dominated by a dense canopy of chromospheric fibrils that covers most of the FOV.
The \halpha\ far wing images at $\pm$40~\kms\ Doppler offset show a large number of elongated absorption features, and a close look at the animation associated with this figure clearly reveals that many display the rapid and complex dynamical evolution that is characteristic for the on-disk counterparts of type-II spicules. 
More interestingly, when comparing the evolution in the red and blue wing images, we see that a majority of the dark threads in the red wing move towards the strong network field concentrations, opposite to the familiar RBE dynamics that is predominantly outward in the blue wing images.  

These structures, which we term as \textit{downflowing} RREs, are described here for the first time and form the major theme of this paper. They are characterized and compared with the traditional RREs and RBEs which have been widely been observed in the recent past. 

\begin{figure*}[ht!]
   \centering
   \includegraphics[width=0.95\textwidth,height=20cm]{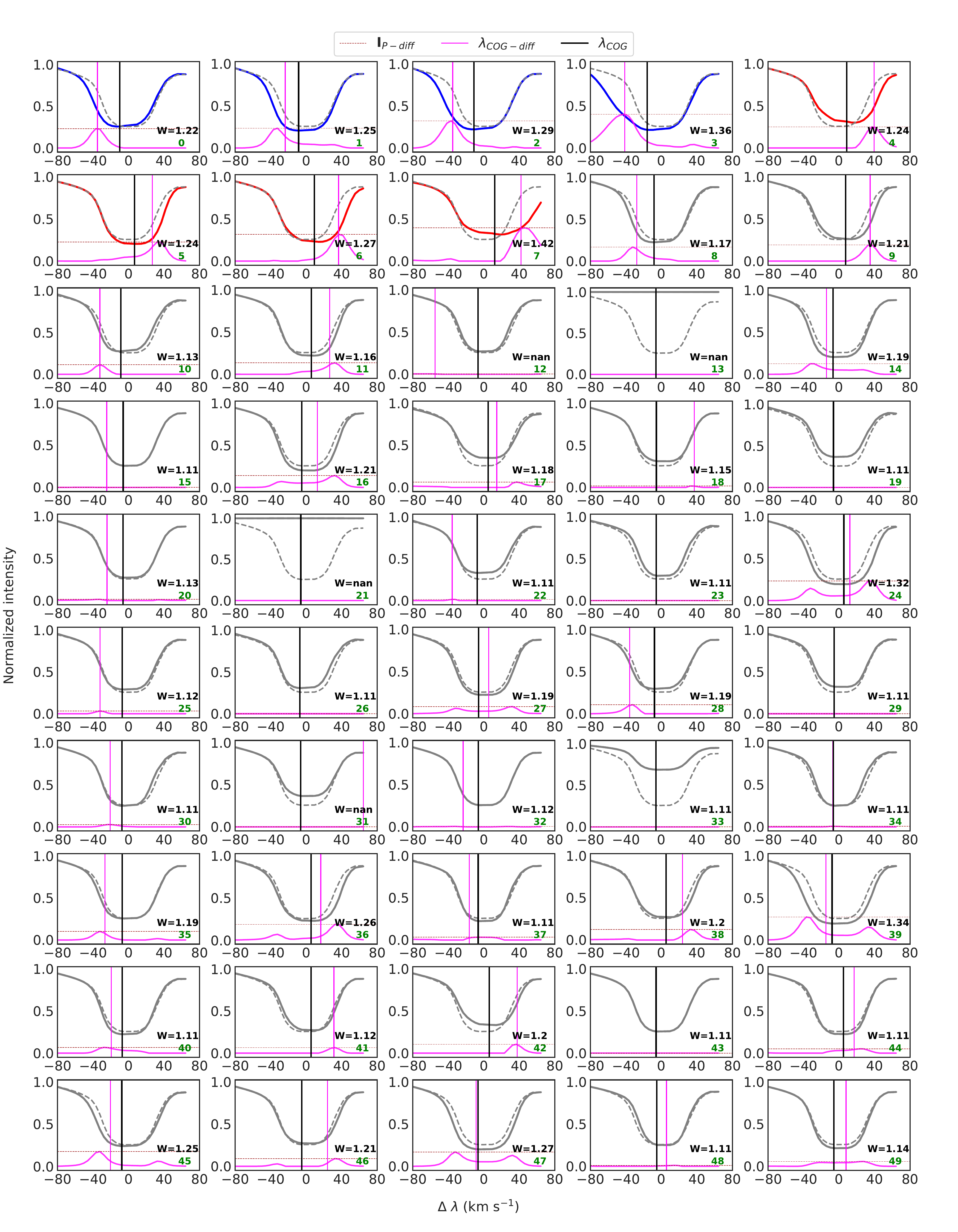}
   \caption{All 50 RPs of \halpha{} spectral lines considered in the final configuration of the $k$-means algorithm. 
   The RP number is indicated in the lower right corner.
   RPs 0--3 are colored in blue signifying blue-ward excursions whereas 4--7 are colored in red indicating red-ward excursions. The remaining RPs (8--49) are drawn in grey. The dashed gray spectral line in each of the panels shows the mean \halpha{} profile averaged over the entire data set and serves as reference. The magenta profiles show the positive part of the difference between the mean \halpha{} profile and the respective RP in each panel. The dashed horizontal magenta line marks the level of the peak (I$_{\mathrm{P-diff}}$) of the differential profiles, and the vertical solid magenta line marks the COG Doppler offset of the positive value of the differential profiles ($\lambda_{\mathrm{COG-diff}}$). The width $W$ of the \halpha{} line core is given for each RP in \AA. 
   }
    \label{figure:cluster_plot_halpha}%
    \end{figure*}
    
\section{Method of analysis}
\label{Section:Method}

\subsection{Characterizing RBEs and RREs/downflowing RREs from their spectral profiles}
\label{Subsection:method-kmeans}

The first and foremost task in our analyses was to identify the rapid blue and red excursions in our data. We followed the $k$-means clustering method 
as described in  
\citet{my_paper_3} 
to identify the different \halpha{} and \cak{} profiles that occur during these rapid excursions. 

The method works by partitioning similar observations into $k$ number of groups or clusters. Each observation is assigned to the cluster with the nearest mean (also called cluster center). It is an iterative algorithm whose main objective is to minimize the sum of distances between the observed points and their respective cluster centers. In our data, each observation (pixel) 
is a spectral profile which is compared with the mean spectral profiles of each of the $k$ clusters before assigning it to one. %
An advantage of using this technique is that it relies on the complete spectral signature of the different features on the FOV rather than their appearance at a particular wavelength position. This enables efficient detection and characterization of the features as discussed below.

Thus, we applied this algorithm to cluster intensity profiles of each pixel on the FOV on the combined \halpha{} and \cak{} data. The detailed algorithmic steps, data pre-processing methods, followed by finding the optimum number of clusters, among which the data can be divided, has been discussed in detail in \citet{my_paper_3}. In this paper, we leverage the analysis performed in \citet{my_paper_3} and proceed with the 50 clusters among which the data had been grouped. Each cluster has a representative profile (RP) that corresponds to the mean over all profiles in that cluster. 
Naturally, each and every pixel in the FOV is uniquely assigned to a particular cluster that can be represented in the form of a 2D map, like the one shown in panel~(c) of Fig.~1 in \citet{my_paper_3} (also see panel~(d) of Fig.~\ref{figure:shadows}).
We base further analyses, performed in this study, solely on the \halpha{} RPs and the quantities extracted from them.
Figure~\ref{figure:cluster_plot_halpha} shows all 50 \halpha{} RPs which resulted from the $k$-means clustering. 
With the exception of RPs-$13$ and $21$, that coincide with the gray borders outside the common CRISP and CHROMIS FOV in Fig.~\ref{figure:Context_v1}, all the remaining RPs correspond to various features observed in the FOV. In the following subsections, we describe the methods undertaken to first identify the RBE and RRE/downflowing RRE RPs and use them further to detect on-disk spicules in our dataset.

\subsubsection{Identifying the RPs}
\label{subsubsection:Identifying-RPs}
RBEs and RREs have typically been observed in \halpha{} and their spectral profiles show significant absorption asymmetries in the blue and red wing positions compared to an average quiet Sun profile \citep{Luc_2009}. These asymmetries are a sign of velocity gradients that are commonly observed in spicules. Preliminary analysis with \verb|CRISPEX| \citep{2012ApJ...750...22V}, a widget-based analysis tool written in Interactive Data Language (IDL), suggests that downflowing RREs have spectral signatures similar to the traditional RREs with an absorption asymmetry in the red wing of \halpha{}. Therefore, the distinction between RREs and downflowing RREs was not made while identifying their characteristic RPs.

We begin our identification strategy by computing the differential profiles, i.e. the difference between average \halpha{} profile and the RP, similar to \citet{Luc_2009} for each cluster. By considering only the positive values of these differential profiles, we have a measure to determine the enhanced absorption part. The positive part of the differential profiles are shown in magenta in Fig.~\ref{figure:cluster_plot_halpha}. This is followed by determining the center-of-gravity~(COG) and the peak intensity of all 
these magenta differential profiles
(their values are indicated by the solid vertical and horizontal dotted magenta lines, respectively). We used the method described in \citet{2003ApJ...592.1225U} for computing the COG, and its position was used as a measure of the Doppler offset of an absorption feature in the RPs.
The combination of large values for COG and positive peak of the difference profile serves as an effective selection criterion to determine enhanced absorption in the \halpha\ wings. 


\citet{2016ApJ...824...65P} highlighted the importance of \halpha{} line width in the detection and analysis of spicules. They reported that the statistical properties measured solely from the \halpha{} line wing images at selected wavelength positions would result in an underestimate of the extracted physical quantities because spicules show a large range of Doppler offsets during their evolution. For such cases line width maps 
provide a more robust tool for spicule detection. Typically, RBEs and RREs show enhanced line widths in addition to large Doppler offsets, with the fastest ones having even broader line profiles. This indicates that both Doppler offset and the line width are important properties that should be considered while detecting spicules. Therefore, we leverage the detection of RBEs, and RREs/downflowing RREs by combining the two factors. We choose the method described in \citet{2009A&A...503..577C} to calculate the widths of the \halpha{} line core over the full-width at half maximum (FWHM) method utilized by \citet{2016ApJ...824...65P}, because FWHM tends to mix the photospheric and the chromospheric signals, which is effectively avoided in the former. Nevertheless, spectral profiles of spicules show enhancements in both FWHM and line core width of \halpha{} rendering the validity of both approaches.

\begin{figure}
  \centering
  \includegraphics[width=0.5\textwidth]{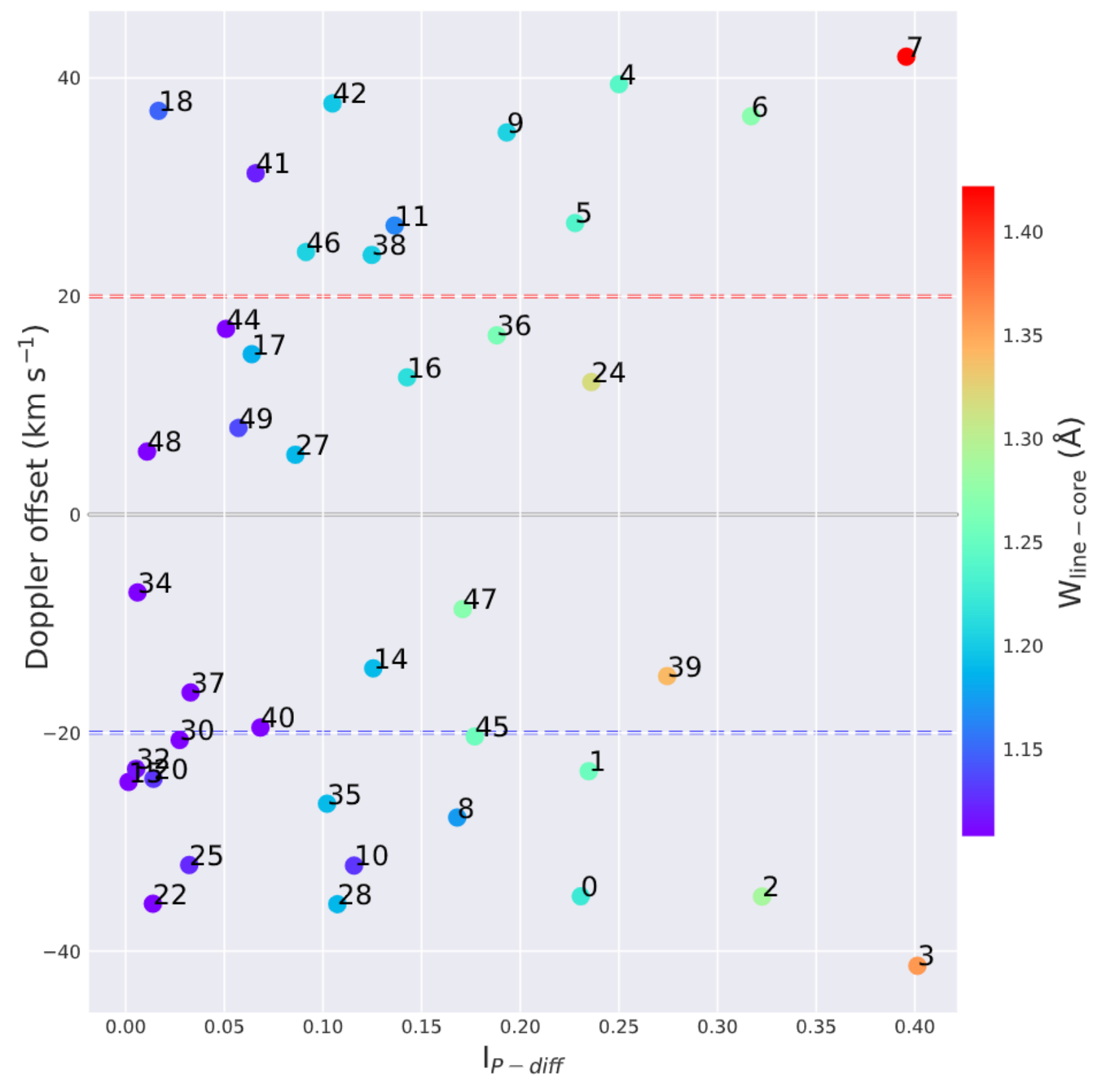}
  \caption{Scatter plot between the COG Doppler offsets and the peak intensity (I$_\mathrm{P-diff}$) of the differential profiles for all RPs. The points are annotated with the respective RP index and color coded based on the \halpha{} line-core width. The horizontal dashed red and blue lines indicate a Doppler offset of $\pm$20~\kms{} as a reference.}
  \label{figure:RP-selection}%
\end{figure}

Figure~\ref{figure:RP-selection} shows a scatter plot between the COG Doppler offsets and peak intensities (I$_\mathrm{P-diff}$) of the magenta difference profiles, the color of the data points follow their respective line core widths. 
Such a representation takes into account all possible spectral characteristics important for RBE/RRE qualification, thereby leading to an efficient and a robust characterization. For visualization purpose, we choose to restrict the lower limit of the colormap to a line core width of $1.1$~\AA. Thereafter, we imposed the following three criteria to consider RPs belonging to RBEs and RREs/downflowing RREs. They should: (1)~have a minimum COG Doppler offset of 20~\kms{}, 
(2)~have I$_\mathrm{P-diff}$ $\geq$ 0.2
and (3)~have line core widths $\geq$ 1.2~\AA. Evidently, RPs $2$, $3$, $6$, and $7$ stand out quite distinctly in Fig.~\ref{figure:RP-selection}, as they not only have the highest Doppler offsets, but also high values of peak intensity and line core widths making them clear RBE and RRE/downflowing RRE RPs. Moreover, the above criteria are also satisfied by RPs-0, 1, 4 and 5, though not as distinctly, whereas the rest of the profiles clearly do not make the cut.
Therefore, we chose RPs 0--7 for the analyses presented in the rest of the paper and assign RPs~0--3 to RBEs and 4--7 to RREs/downflowing RREs in increasing order of the strength of their spectral properties and they are accordingly arranged in the first 7 panels of Fig.~\ref{figure:cluster_plot_halpha}. The remaining RPs, numbered from 8--49, are indicated in gray. 

According to Fig.~\ref{figure:RP-selection}, the computed Doppler offsets of RP-0 and RP-4 are comparable to RPs-2 and 6 respectively, but, the latter show stronger absorption in their line wings and a stronger shift in their spectral lines that in turn also contributes to an increased line core width making them stronger than the rest. Therefore, RPs-0 and 4 are assigned lower in the order of labeling the RPs in Fig.~\ref{figure:cluster_plot_halpha}. Additionally, we also exploited the difference between the spectral properties of RPs~6 and 7 (2 and 3) and RPs~4 and 5 (0 and 1) described earlier, and segregated the detected RBEs and RREs/downflowing RREs into two compartments with the purpose of comparing them. Basically, we grouped RPs~6 and 7 (2 and 3) for the red shifted (blue shifted) events under stronger red (blue) excursions, whereas RPs~4 and 5 (0 and 1) were grouped together under weaker red (blue) excursions in a way such that the events in one group were unique with respect to the other. Such a segregation formed the basis of the investigations carried out in Sects.~\ref{subsection:DRREs-observation} and \ref{Subsection:spatial-distribution-spicules}, where we discuss them further. In addition to the above properties, we also note from Fig.~\ref{figure:RP-selection} that the line-core widths of downflowing RRE/RRE-like RPs are slightly enhanced in comparison to the RBE-like RPs. The difference is more pronounced for the RPs belonging to the strong excursions with respect to the weaker ones. We would like to note that the estimates based on the COG technique serves well to identify RBE/RRE like profiles, but it fails for some of the other RPs (such as RP-12 and 31) for which this technique is rather meaningless. 
Nevertheless, the method adopted in this paper provides one of the most comprehensive approach so far in characterizing spicular spectral profiles.

\begin{figure}
  \centering
  \includegraphics[width=0.49\textwidth]{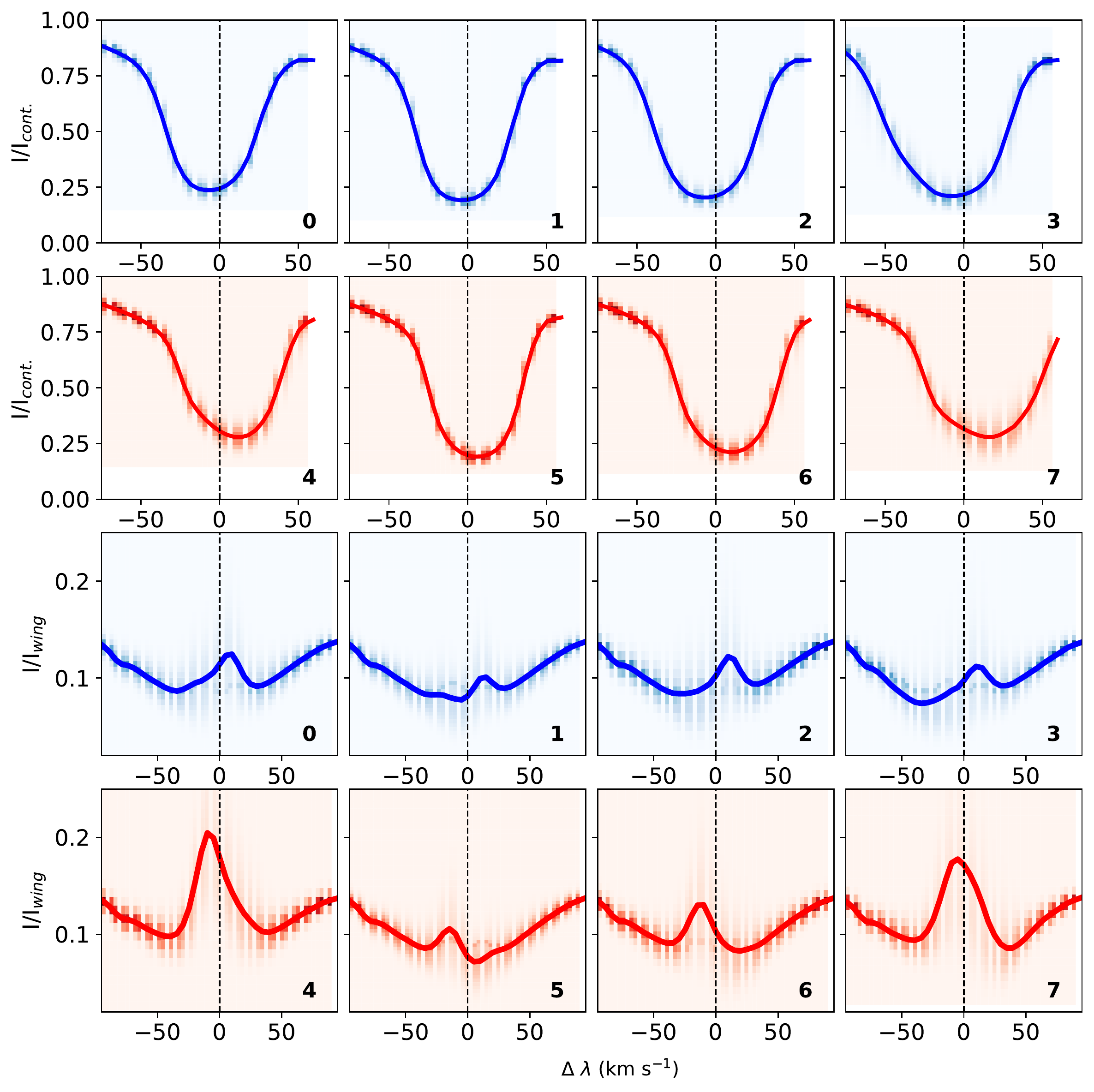}
  \caption{Density plots of the \halpha{} (top two rows) and \cak{} (bottom two rows) spectra for RBE and RRE/downflowing RRE RPs (0--7). The density (darker meaning higher concentration of spectra) corresponding to each RPs shows the distribution of profiles over the whole time series. The solid lines overplotted correspond to the RPs discussed in the text. The blue (red) color coding indicates the excursions in the blue ward (red ward) side of \halpha\, respectively.
  }
  \label{figure:1d_density_plots_H}%
\end{figure}

Now, once the RPs have been identified, a question arises as to how well are they able to represent individual profiles in a particular cluster? Figure~\ref{figure:1d_density_plots_H} sheds light in this direction and shows the RBE and RRE/downflowing RRE-like spectral profiles in \halpha{} (top two rows) and \cak{} (bottom two rows) for each RP in the form of density distributions, with darker shades indicating a higher number density of spectral lines. The colored solid line in the different panels shows the mean over all profiles for a particular cluster, which, in our case, is equivalent to an RP. We clearly see that the distribution of the spectral profiles belonging to each cluster is narrow and is mostly concentrated near their respective mean profiles. This indicates that the identified RPs are able to efficiently describe the profiles in the respective clusters for both \halpha\ and \cak{}.

From the distribution of the spectral profiles in \cak{}, we see that for a vast majority of cases, the strongest Doppler shifted K$_{3}$ (in both the red and blue excursions) has a significantly stronger opposite K$_{2}$ intensity enhancement relative to its line core. In other words, we see that the absolute difference between the intensities $I$(K$_{2}$)~-~$I$(K$_{3}$) is correlated with the shift of K$_{3}$, as was also shown in \citet{my_paper_3}. This is the case exclusively for the RBE and RRE-like RPs, which also formed an additional basis for the identification of the RPs in \cak{}. The rest of the profiles show no such characteristic behavior \citep[refer to Fig. B.2 in the appendix of][]{my_paper_3}. 

The intensity enhancement in the K$_{2}$ peaks is inherently linked to lower layer photons,  observed due to the presence of strong velocity gradients in spicules which remove top-layer opacities at those wavelengths. Such lower layers probably feature enhanced emission as it is due to increased temperature or other local source function enhancing effects, but the relation between the K$_{2}$ and the  K$_{3}$ features, set by velocity and reproducible by a simple model, reveals the role of the velocity gradients. In \citet{my_paper_3} this was described as an opacity window effect due to the two-dimensional presence of background features imaged in K$_{2}$ wavelengths and to distinguish it from other effects such as the reflector effect \citep{1981ApJ...249..720S,Scharmer_1984}. Similar enhancements are generally observed whenever there is a gradient in the LOS velocity in strong resonant lines such as \cak{} or \Mgk{} \citep{Carlsson_1997,2015A&A...573A..40D,2015ApJ...813..125K}.
A further example in a very different solar feature is found in umbral flashes in the \Mgk{} spectral line \citep{Bose_2019}. For a recent discussion on this topic see Sect.~4.1 of \citet{2020A&A...642A.215H}. 

\subsection{Spicule detection and Morphological operations}
\label{Subsection:method-morphology}

\subsubsection{Halos around spicules and their substructures}
\label{Subsection:shadows}

\begin{figure*}
   \centering
   \includegraphics[width=\textwidth]{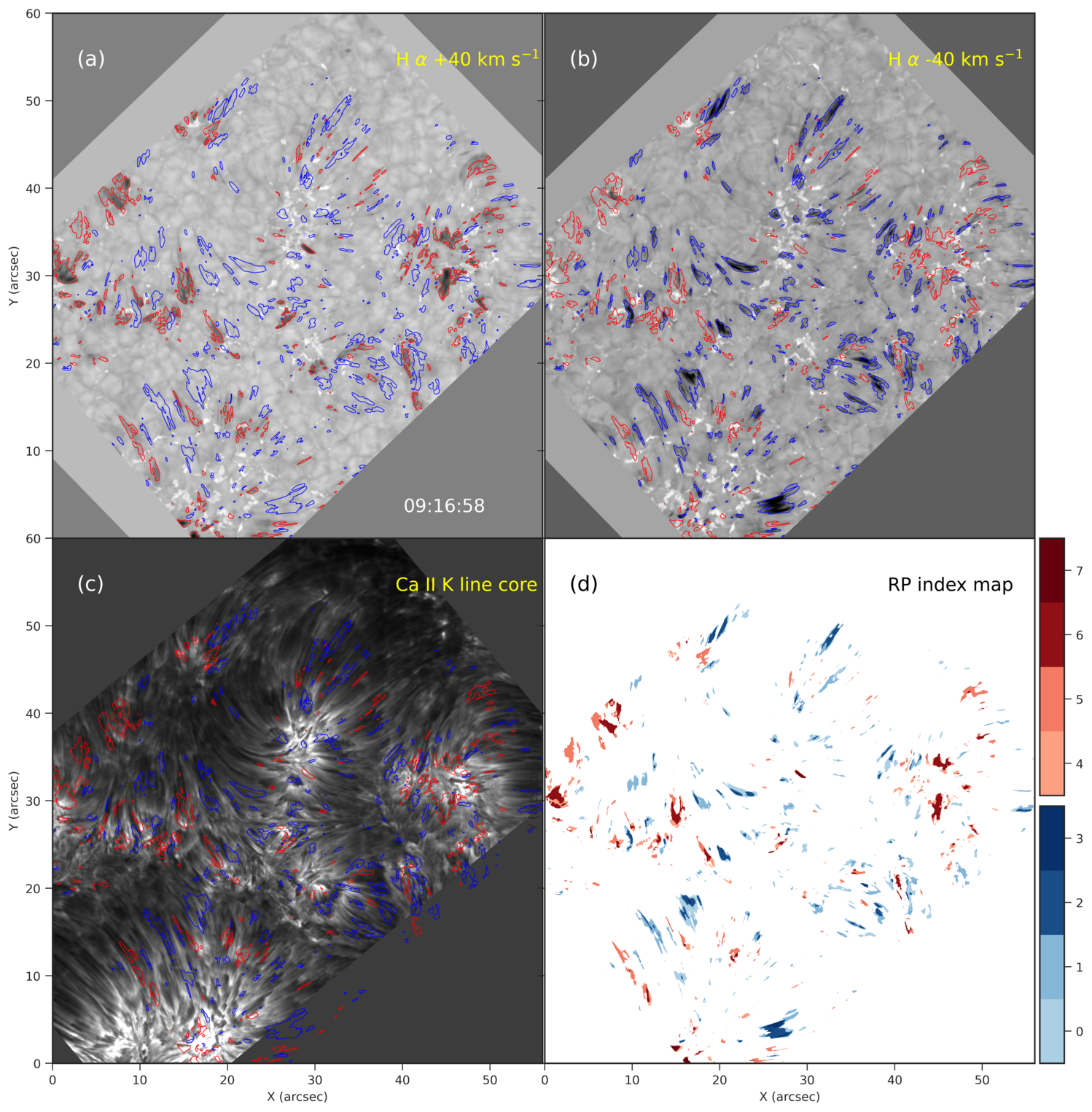}
   \caption{Overview of the detected RBEs and RREs/downflowing RREs in \halpha{} wing and \cak{}. Panels~(a), (b) and (c) show RBEs and their associated substructures (halos) in blue and RREs/downflowing RREs and their halos in red colored contours against a background of \halpha{} wing images at +40~\kms{}, $-$40~\kms{} and \cak{} line core, respectively at time t=09:17UT. Panel~(d) shows the corresponding RP index map with gradients in the color indicating the contribution from the different RPs as shown in the colorbar. An animation of this figure is available at \url{https://www.dropbox.com/s/wav35xo38vo6s9d/shadow_movie1.mp4?dl=0}.
   }
    \label{figure:shadows}%
    \end{figure*}

\begin{figure*}
   \centering
   \includegraphics[width=1\textwidth,trim={0cm 0cm 0cm 0cm},clip]{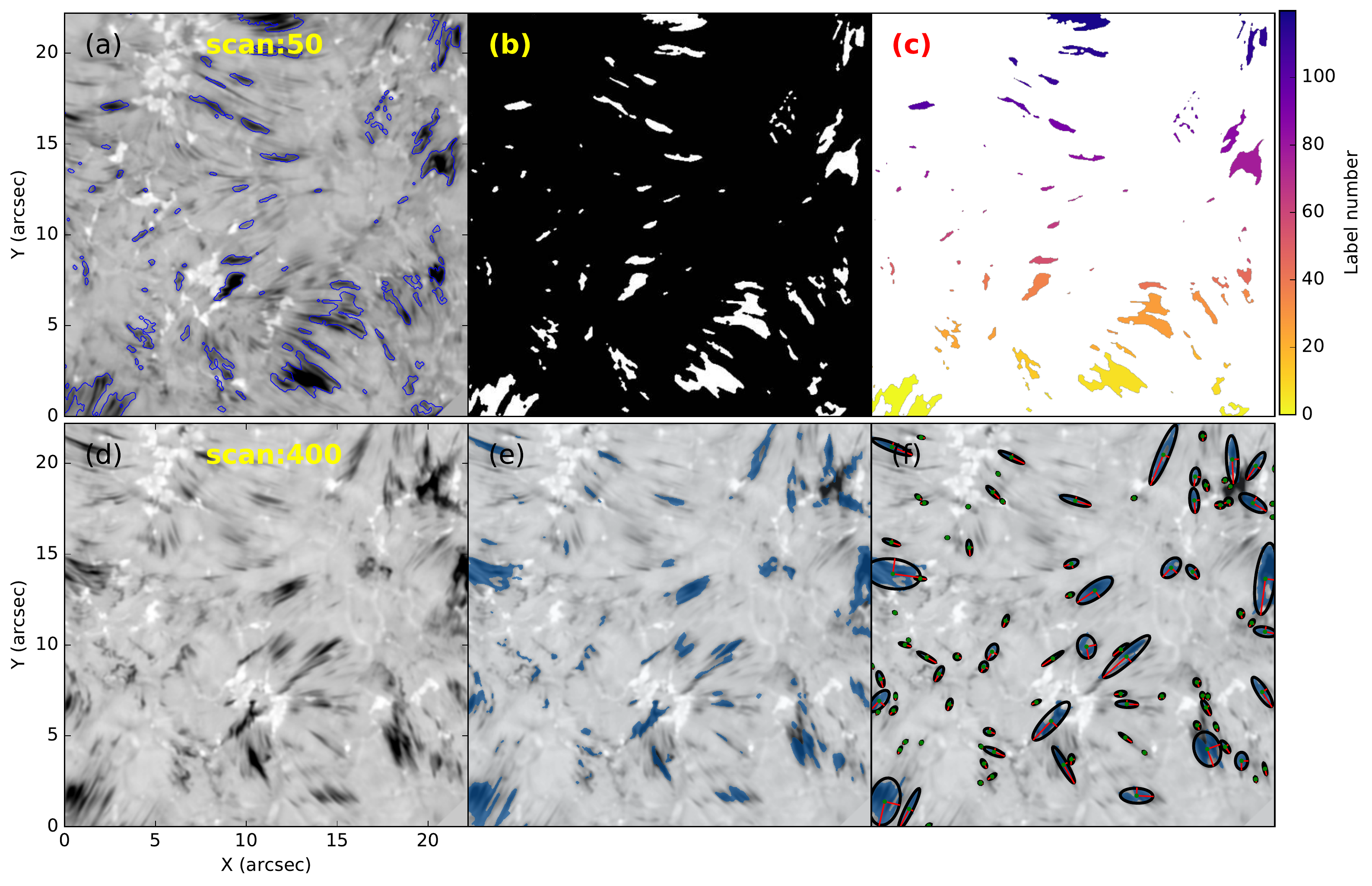}
   \caption{Overview of the morphological processing techniques described in the text. Panel~(a) shows a portion of the FOV in \halpha\ wing ($-38$~\kms) at time t=09:24~UT (scan:50). Blue contours on this image indicate the RBEs detected with RPs 0--3 as discussed in the text, panel~(b) shows the binary mask of the detected RBEs, and panel~(c) indicates the result after performing connected component labeling of the binary mask in panel~(b). The labels are color coded as seen in the adjoining colorbar. The total number of labels for RBEs over the full FOV and full duration corresponds to 19,643. Panel (d) shows the same FOV as in (a) but at time t=10:43~UT (scan:400). Cobalt blue colored contours in panel (e) show the masks of detected RBEs obtained after performing morphological labeling, and (f) shows the fitted ellipses with their centers (green), semi-major axis and semi-minor axis (red) for each label shown in (e).
   }
    \label{figure:sample_detection}%
    \end{figure*}



Spicules are known to display complex dynamical behavior owing to their: (a) flows aligned along the chromospheric magnetic field lines, (b) swaying motions and (c) torsional motions \citep{Bart_3_motions}. Furthermore, \cite{2013ApJ...769...44S} established that many RBEs and RREs exhibit a variation in Doppler offsets along their lengths and breadths due to which the widths of \halpha\ spectral lines are enhanced, as was shown by \citet{2016ApJ...824...65P}. This, in turn is reflected in their spectra which essentially translates to having multiple spectral profiles (RPs) for the same spicular body. 
In this section, we attempt to fully capture this complex behavior by showing how the different RPs contribute in detecting the complete morphology of spicules (including the downflowing RREs) and their implications.

Figure~\ref{figure:shadows} and its associated animation shows an overview of the spicules detected by including multiple RPs (0--7). Panels~(a) and (b) respectively show the contours of the detected RBEs and RREs/downflowing RREs in blue and red colors against \halpha{} +40~\kms{} and $-$40~\kms{} images. A close look at panel~(b), for example, clearly reveals that in many cases the overlaid contours enclose a region that appears to be larger than the visible dark thread like structures, such as the ones around ($\mathrm{X}$,$\mathrm{Y}$) = (27\arcsec,5\arcsec) or (15\arcsec,17\arcsec). Similar examples can also be found in panel~(a) for the RREs/downflowing RREs. These lighter shades around the central dark regions are here termed as spicule halos. They represent structures with weaker Doppler offsets surrounding the centrally stronger offset regions. The halos become prominent in the images observed closer to the \halpha{} core, implying that though they have weaker Doppler offsets they form a part of the same morphology and evolve together. 
Panel~(c) shows the same contours as in (a) and (b) but against a background of a thick chromospheric canopy imaged in \cak{} line core. As shown in \citet{my_paper_3}, spicules do not have a sharp intensity contrast in \cak{} (due to the opacity window effect) unlike \halpha{}, which makes it nearly impossible to observe them as we do in the \halpha{} wing images. However, looking at the animation of this panel and closely following the overlaid contours does provide a better impression of the evolution of spicules in such band-pass. 

Further justification of including multiple RPs in the analysis and visualization become clearer when we look at the RP index map shown in panel~(d) of Fig.~\ref{figure:shadows}. It shows the combined RBE and RRE/downflowing RRE RP indices with varying strengths as set by their their associated Doppler offsets, line core widths and peak intensities of the respective enhanced differential profiles. The darker blue (red) shades are an indicator of higher values of the above three parameters compared to the lighter ones. Clearly, we see that in several cases the darker shades are accompanied or surrounded by lighter shades or the previously described halos, in both blue and red shifted structures. The animation of this figure distinctly indicates that these halos evolve in conjunction with their corresponding darker cores making it clear they are a part of the same, but bigger, morphological structure. As seen earlier for panels~(a) and (b), this again strongly indicates widespread group behavior among spicules. Such behavior was also pointed out by \citet{2014ApJ...795L..23S} among the off-limb spicules. The animation also shows a large number of downflowing RREs (in red color), co-evolving with their halos, with an opposite apparent motion compared to the RBEs/RREs (in blue/red color). 

The discussion presented in the preceding paragraphs demonstrates that spicules have multiple substructures that are possible to be detected only after the inclusion of multiple RPs. It also further evinces that it is inaccurate to infer about their nature based on detections from a single wavelength position since their morphology, length and lifetimes could also be very different in reality. As an example, we show the multi-structural features among the red and blue excursions belonging to the stronger category in Appendix~\ref{Appendix:substructures}. Such a representation makes the variation in their spectral properties abundantly clear and prominent.

\subsubsection{Morphological processing techniques}
\label{Subsubsection:RBE_RRE_detection}
The discussion presented in the preceding section justifies the importance of including multiple RPs in spicule detection. This section further advances and describes the details of the image processing algorithms employed in detecting the RBEs and RREs/downflowing RREs. The remainder of this section describes the detailed steps followed in the detection of RBEs in our data. The exact same procedure was also employed in the detection of RREs/downflowing RREs. 

We started off by creating a 3D binary mask (in spatial and temporal domains) containing all the pixels in the FOV belonging to RPs~0--3 and assigning them a value of 1 (bright) and the rest 0 (dark). A morphological opening, followed by a closing operation, with a 3$\times$3 diamond shaped structuring element was applied to each of the binary masks on a per time step 
basis. The opening operation is analogous to an erosion followed by a dilation, and is useful to remove tiny bright dots in the binary mask. This helps to get rid of small (1-pixel) connectivity that might be present between different morphological structures in the 2D space. A morphological opening operation however, can also create small dark holes in between bright structures which can be effectively closed by using a closing operation that is the reverse of opening. It is however important to keep the same 3$\times$3 structuring element as before. These operations were performed on a per time step 
basis, i.e. in 2D, because we intended to preserve the connections in the temporal domain that would enable us to effectively label them.

The next step was to perform a 3D connected component labeling \citep{labeling_1996} so that we could identify components uniquely based on a given heuristic. This technique is widely used in computer vision and image processing technology. Two pixels are said to be connected when they are neighbors and have the same numerical value. In this case, we aim to label pixels in 3D space that are connected and are similar in spectra as set by the $k$-mean procedure. To not bias for direction we selected an 8-neighbourhood connectivity in 3D. 
The top row of Fig.~\ref{figure:sample_detection} (panels~(a)--(c)) provides an overview of the steps undertaken for the detection of RBEs at the indicated scan.  

When applied to the complete dataset, the method described above led us to identify 19,643 RBEs and 14,650 RREs (including the downflowing RREs). The two panels of Fig.~\ref{figure:appendix-spicules} in Appendix~\ref{appendix-supp-figs}, shows the location of all the detected red and blue excursions respectively, over the co-spatial CRISP and CHROMIS FOV for the entire time series. On average, most of the spicular activities are seen to exist in the vicinity of the strong field regions or the network fields (shown in black colored contours).
The difference between the number of RBE and RRE/downflowing RRE detection is consistent with \cite{2013ApJ...769...44S}, with the RRE:RBE detection ratio being <~1. However, it is important to remark that our red excursions also consists of the downflowing RREs along with the traditional RREs, and we report far more number of on-disk red and blue excursions than any of the preceding works cited in this paper, mainly because of the unique detection method followed by advanced morphological operations discussed above. Therefore, this allowed us to perform much more exhaustive data analysis than many such earlier works. 

\subsection{Dimensional analyses and lifetime statistics}
\label{Subsection:method-stats}
One of the major goals of this study is to statistically compare the properties of the newly reported downflowing RREs to the traditionally known RBEs and RREs. The detection method undertaken in Sect.~\ref{Subsection:method-morphology} yielded a large number of on-disk spicules which provides a perfectly fertile ground to explore further in this direction. In this section, we focus on the technique employed to compute their dimensions and lifetimes.

We began by fitting an ellipse to each of the labels obtained after the 3D connected component labeling, and then computing their lengths of the major axis, eccentricities ($e$, that is given by $e=\sqrt{1-{b^2}/{a^2}}$,
with $a$ and $b$ being the semi-major and semi-minor axis of a standard ellipse, respectively),  and the occupied area. Panels~(e) and (f) of Fig.~\ref{figure:sample_detection} provide an illustrative example where we show the identified spicules in cobalt blue contours for a given scan (shown in panel~(d)), along with their fitted ellipses--together with their centers and semi-major axis, for a small area on the FOV. To avoid erroneous detections, a lower limit for the length of the major axis of detected blue shifted and red shifted features was set at \textasciitilde100 km or 4 CHROMIS pixels, such that any label with length below the threshold is not included. Strictly speaking, spicules are not perfect ellipses. However, they share a common morphology where their lengths (generally) far exceed their widths \citep{1968SoPh....3..367B,Tiago_2012}, making them appear as dark elongated streaks when observed in \halpha{} on-disk wing images. The elliptical fitting is performed solely for the purpose of determining the length occupied by RBEs/RREs by measuring the length of their respective major axes which ensures that the ends are located at the widest points of the perimeter of the rapid excursions. Furthermore, elliptical shapes allow a certain degree of freedom even for those spicules that are not highly elongated but rather have their lengths only slightly greater than their widths. These shapes can readily be fitted by ellipses with $0\leq e \leq 1$, which would mean that the length of the major axis of the ellipses provide a good approximation of the length of RREs/downflowing RREs and RBEs.

Earlier studies, such as \citet{Luc_2009} and \citet{Sekse_2012,2013ApJ...764..164S}, relied on obtaining the morphological skeleton of the detected RBEs and RREs in order to compute their lengths. In such cases, the skeleton is a thin version of that shape that is equidistant to its boundaries. The major axis of an ellipse is very similar to the skeleton because it passes through the COG of the ellipse and therefore their lengths will be comparable. Both these techniques are basically approximations and skeletonizing certainly has its own benefits as it preserves the shape of any given structure. However, since we are interested in determining the maximum extent from both ends of a feature, the major axis of an ellipse can prove to be advantageous for complex morphological structures that are often associated with RBEs and RREs.


It is however important to recall that the 3D connected component labeling produces a chain of events that are attached both in space and time. Therefore, if an event (or a label) lasts for multiple time frames, we only consider the length, area and $e$ when it is at its maximum extent. A similar approach is followed for spicules with multiple structures or halos around them. This is justified because inclusion of multiple dimensions for the same spicule would lead to incorrect statistics. The results obtained after performing the analysis are shown and discussed in Sect.~\ref{Subsection:results-dimensions}.

The lifetimes of spicules were determined on a per event basis, where we considered the difference between the first and the last occurrence of the same event in the temporal dimension. The lower limit of the measured lifetimes is set by the CRISP cadence of \textasciitilde 19.6 s. The results are further discussed in the latter half of Sect.~\ref{Subsection:results-dimensions}.

\subsection{Apparent motion of spicules}
\label{subsection:method-tracking}

Evidently, the major difference between a downflowing and a traditional RRE is their plane-of-sky (apparent) motion with respect to the strong magnetic network areas. In this paper, we investigate the apparent motion of 19,643 RBEs and 14,650 RREs/downflowing RREs on a morphological event-by-event basis, so as to statistically analyze and describe their trajectories in the plane-of-sky. Such an analysis would help us to get an idea as to what extent the red shifted excursions show an opposite trajectory in our datatset, thereby revealing the abundance of such events. We computed the area weighted COG of each morphological label (spicule) and followed their evolution in both space and time. The weighting by area allows the algorithm to follow the trajectory of the larger substructures in a label. It starts tracking the COG from the first occurrence of a label and continues until the last. The $X$ and $Y$ coordinates of the COG were stored for each time step and and were then plotted to display their apparent motion. The results are described further in Sect.~\ref{subsection:DRREs-observation}.

\section{Results}
\label{Section:results}

\subsection{Downflowing rapid red shifted excursions}
\label{subsection:DRREs-observation}

\begin{figure*}[htb!]
   \centering
   \includegraphics[width=0.76\textwidth,keepaspectratio]{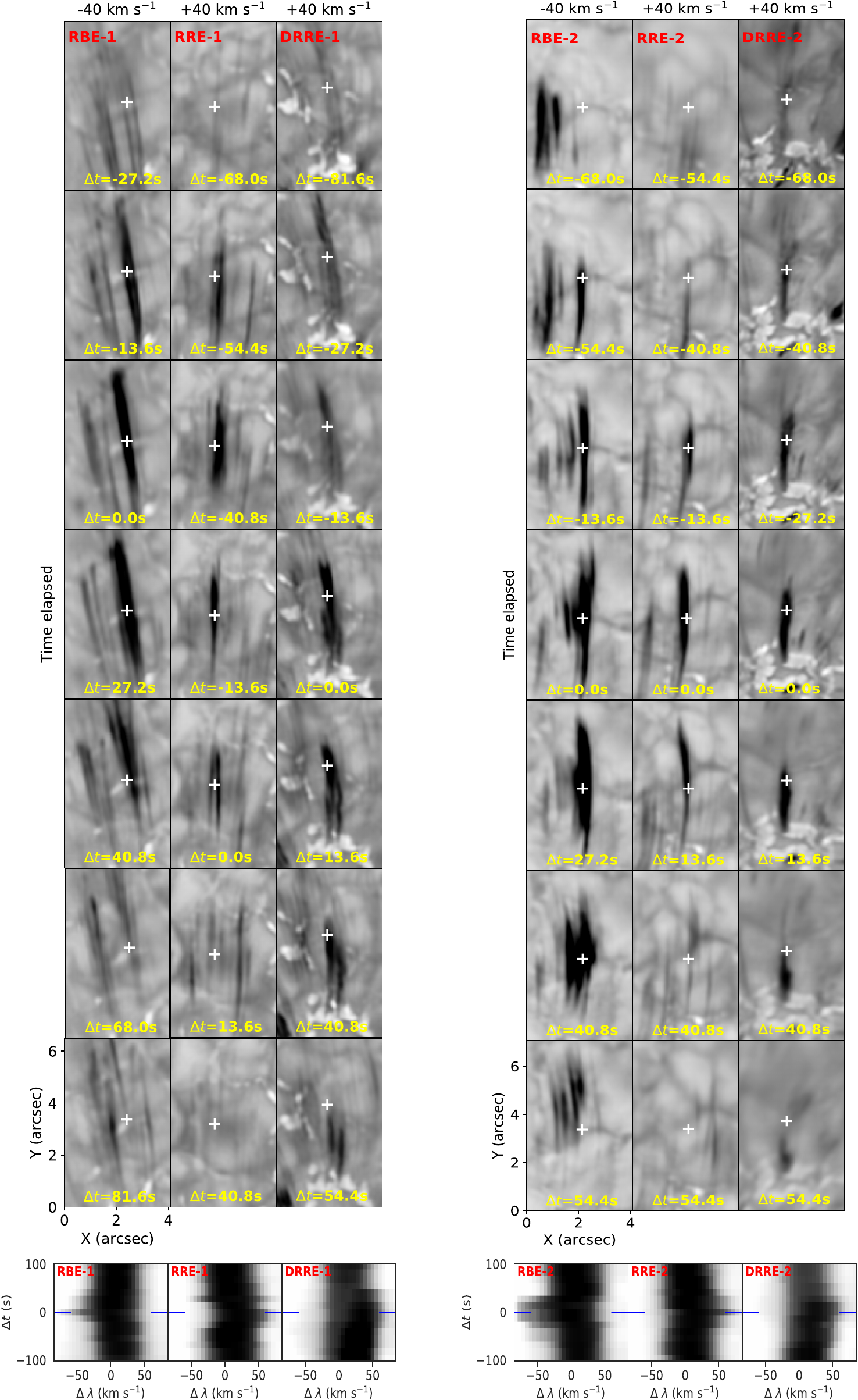}
   \caption{Examples of RBEs, RREs and downflowing RREs (labelled as DRREs). The top panels show their spatio-temporal evolution in \halpha{} blue and red wing images, whereas the bottom row shows the spectral evolution in  $\lambda t$-diagrams corresponding to the locations marked with white crosses. Animations are available at \url{https://www.dropbox.com/sh/3x3kvx158u57xho/AAAA5Xm0sH8RSxPyxe8vqA5Wa?dl=0}
   }
    \label{figure:DRREs}%
    \end{figure*}
     

\begin{figure*}[ht!]
   \centering
   \includegraphics[width=\textwidth]{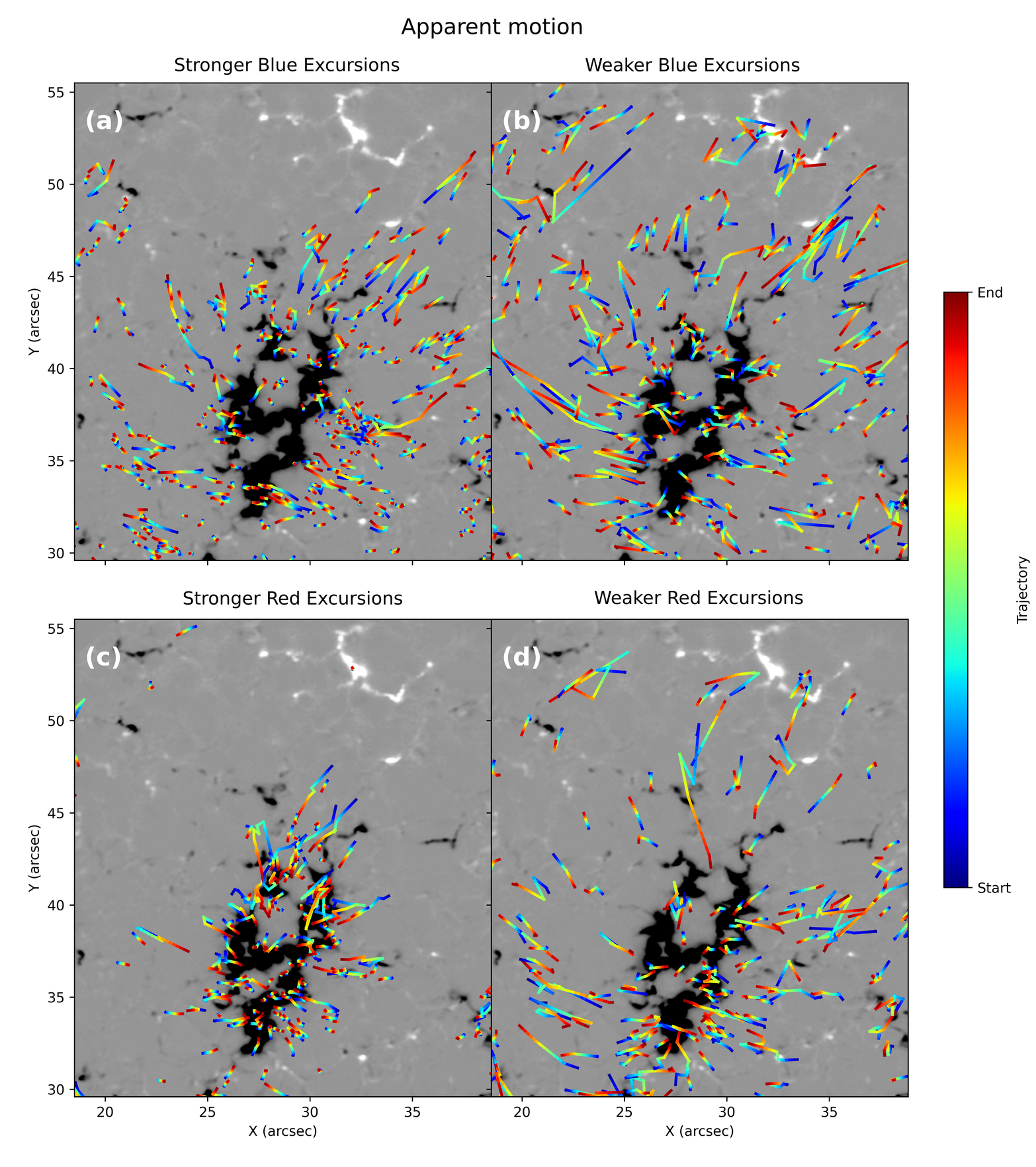}
   \caption{Plane-of-sky trajectories of RBEs, RREs and downflowing RREs mapped according to the strength of their excursions towards the blue (top row) and red (bottom row) wings. 
   The background shows a B$_{\mathrm{LOS}}$ map  saturated between $\pm$ 500~G.
   The rainbow colored trajectories map the travel direction from the origin in blue to the termination in red. 
   Here, only a zoom in on a network patch is shown, Fig.~\ref{figure:appendix-full-fov-apparent-motion} shows the full FOV.
   }
    \label{figure:zoom-in-apparent-motion}%
    \end{figure*}
The animations linked to the \halpha\ red wing images in Figs.~\ref{figure:Context_v1} and \ref{figure:shadows} seem to convey the impression that in many cases the apparent motion of the events is rather inward moving instead of the characteristic outward movement (with respect to the background network areas) associated with traditional RREs. Except for their apparent motion, we find that these events are very similar in their appearance and morphology of RBEs and RREs. In this section, we report clear observational and statistical evidences of these new class of events termed as downflowing RREs. 

Figure~\ref{figure:DRREs} shows two examples each of RBEs, RREs and downflowing RREs (labelled DRREs in the figures). 
The temporal evolution of these events are shown along the column for each case
and they immediately suggest that, morphologically, the downflowing RREs are very similar to both RBEs and RREs, and like RREs they appear in the far red wings of the \halpha{} line core at +40~\kms{}. However unlike the former, the downflowing RREs clearly move towards the bright network structures as they evolve, whereas the traditional RBEs and RREs move outwards in the opposite direction. The online animation associated with each of the examples shown in Fig.~\ref{figure:DRREs} establishes the scenario quite convincingly.

The temporal evolution of their corresponding \halpha{} spectra shown in $\lambda t$-diagrams 
also display typical type-II spicule-like behavior for the downflowing RREs, 
with a sudden development of a highly asymmetric line profile towards the red side of the line core. This is very similar to the characteristic RRE $\lambda t$ evolution first reported by \citet{2013ApJ...769...44S}. For the sake of completeness, we also show extended $\lambda t$-diagrams in Appendix~\ref{figure:appendix-lambda-t-extended} for the two downflowing RREs stretching well before (\textasciitilde 550~s) the occurrence of the red excursions. They clearly show no signs of preceding blue shift that are typical for type-I spicules. Moreover, we also note that the lifetime of their excursions in the red wing are similar to RREs and RBEs. Furthermore, the redshifts associated with the downflowing RREs, in addition to their inward apparent motion, strongly suggests that these are real plasma flows (moving away from the observer) and not simply apparent motions that are often associated with the TR network jets \citep{2014Sci...346A.315T}. The high apparent speeds in the network jets (of the order of 100--300 \kms{}) are most likely not caused by real mass flows. Instead, they are most likely due to heating fronts propagating at Alfvenic speeds \citep{2017ApJ...849L...7D}. Therefore, appertaining to their spectral and morphological similarities, we suggest that the downflowing RREs are basically like RREs simply with an opposite plane of sky motion. 

To rule out the possibility that these are singular events, we performed a statistical study based on the COG tracking method described in Sect.~\ref{subsection:method-tracking}, where each and every event belonging to the blue and red side of the \halpha{} line core was tracked individually. Both the blue and red excursions were first segregated into two compartments, according to the strength of their spectral features, by grouping them in the manner described in Sect.~\ref{subsubsection:Identifying-RPs}. Consequently, after morphological processing operations, events that have at least one stronger RP (belonging to either 2 and 3 or 6 and 7) over their whole lifetime are considered under the stronger excursion category, whereas the rest are grouped into weaker excursions. Panels~(a)--(d) of Figs.~\ref{figure:zoom-in-apparent-motion} and \ref{figure:appendix-full-fov-apparent-motion} show the apparent trajectories of these excursions as indicated in their title.
Figure~\ref{figure:zoom-in-apparent-motion} shows a zoom in to the central network patch and Fig.~\ref{figure:appendix-full-fov-apparent-motion} shows the full FOV. 
The direction of motion is shown by drawing the trajectories of the COG in a rainbow colormap, where the blue marks the origin and red the final destination.
For the sake of clarity, we display only those events that have sufficient apparent displacement: $\geq$~0\farcs5. 
($\geq$~1\arcsec for the full FOV in Fig.~\ref{figure:appendix-full-fov-apparent-motion}). 
The $B_\mathrm{LOS}$ map is shown as background
in order to enhance the visibility of the trajectories and to facilitate better understanding of the apparent motion of the events with respect to the strong magnetic field regions. 

We immediately notice that almost all the events in panels~(a) and (b) originate close to the network region and move further outwards as they evolve--a behavior that is typical to RBEs. The bottom row, on the other hand, shows that the paths traversed by the red excursions are predominantly opposite to their blue shifted counterparts. Panel~(c), for example, shows an inward apparent motion for all the red excursions. Moreover, the origin of the majority of such events can be seen to lie outside the enhanced magnetic field region and they tend to terminate within or in close proximity to the boundary of the strong network regions. The weaker red excursions in panel~(d) mostly shows a mixed-bag scenario where a large number of events show the inward apparent motion, but many among them trace the traditional RRE trajectories, such as the ones around ($X$,$Y$) = (22\arcsec,32\arcsec) or (35\arcsec,45\arcsec). A similar trait is also observed for a vast majority of events detected in the full FOV, shown in panels~(c) and (d) of Fig.~\ref{figure:appendix-full-fov-apparent-motion}, which strongly suggests that the newly reported downflowing RREs are widely prevalent in our dataset in conjunction with RBEs and RREs.

\subsection{Spatial distribution of the stronger and weaker excursions}
\label{Subsection:spatial-distribution-spicules}
\begin{figure*}[ht!]
  \centering
  \includegraphics[width=\textwidth]{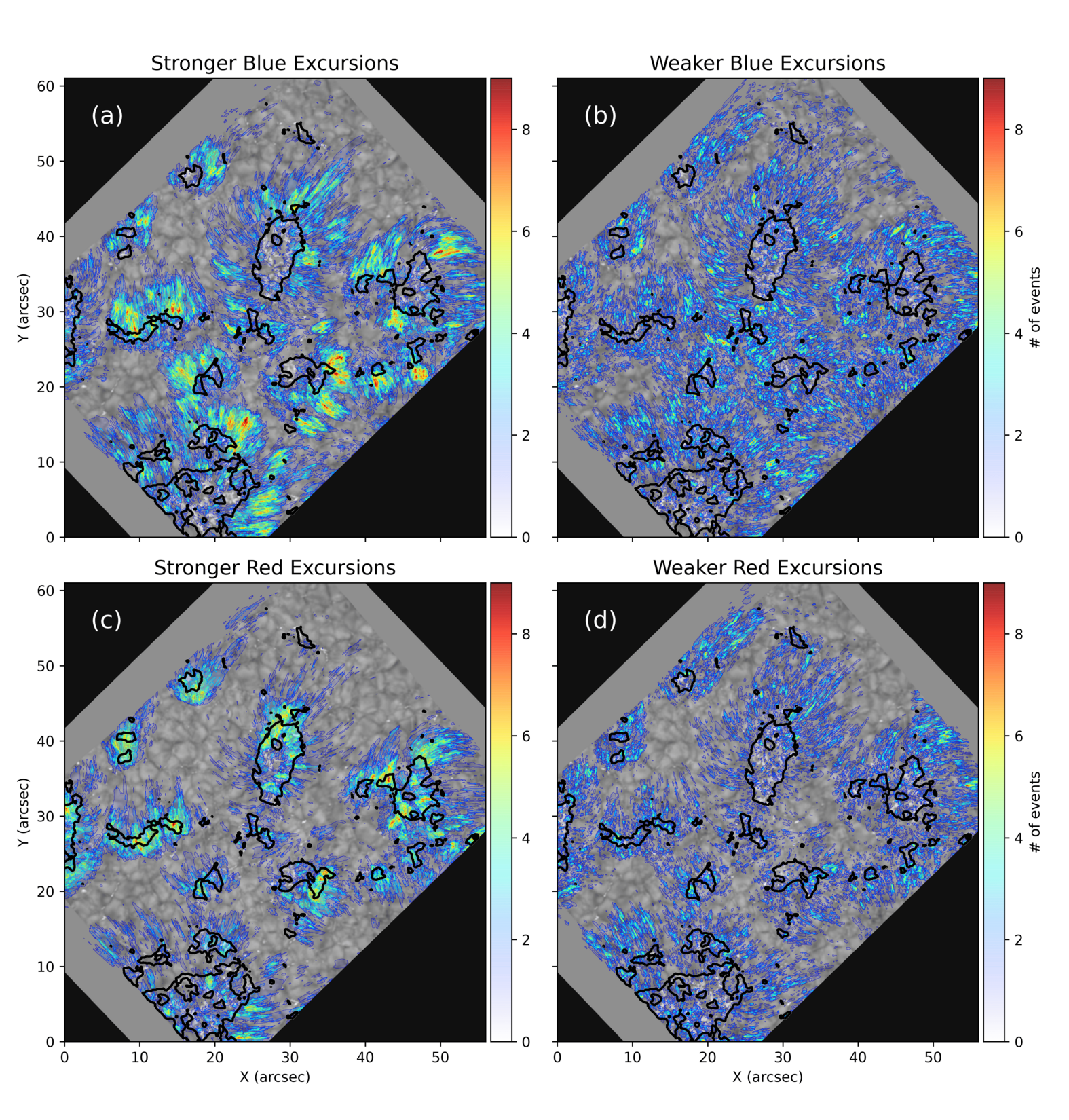}
  \caption{Spatial occurrence of the stronger and weaker blue and red excursions with respect to the strong field network regions. Panels~(a) and (b) show the stronger and weaker blue shifted excursions (RBEs) whereas panels~(c) and (d) shows the distribution of the stronger and weaker red shifted excursions (RREs/downflowing RREs), respectively. The colors represent the number density of the events shown. The events are mutually exclusive, meaning that the excursions shown in any one panel are unique and are not related to the events in the other. The black contour indicates the regions with an absolute LOS magnetic field $\geq$ 100~G.
  }
    \label{figure:unique_spicules_velocity}%
    \end{figure*}

The trajectories of the strong and weak excursions presented in the preceding section sparks interest in investigating their detailed spatial occurrences with respect to the background network areas over the full FOV. Traditionally, RBEs and RREs are known to appear in the close vicinity of strong magnetic field network regions which are also thought to be their foot-points \citep{Luc_2009}. Therefore, it is worthwhile to explore if such a behavior is also seen among the stronger and weaker excursions detected in this study.
Figure~\ref{figure:unique_spicules_velocity} shows the occurrence of these excursions in the form of 2D density maps. Once again, the events were grouped in exactly the same way as in Sect.~\ref{subsection:DRREs-observation}, with the stronger excursions belonging to the group with RPs~2 and 3 (6 and 7), shown in panels~(a) and (c), and the weaker ones belonging to RPs~0 and 1 (4 and 5) panels~(b) and (d), in such a way that events belonging to these categories are mutually exclusive with respect to one another. 

A closer examination of Fig.~\ref{figure:unique_spicules_velocity} reveals that stronger excursions are located closest to the enhanced network regions (indicated by black contours), whereas their weaker category counterparts are located further outwards. Moreover, panels~(b) and (d) also suggest that the weaker spicular excursions appear to exist all over the FOV, but their number density is, on average, roughly 10--15\% lesser than their stronger counterparts. Panels~(a) and (c) also highlight an important difference between the stronger blue and red excursions. In panel~(a) it appears that the density of blue excursions are mostly concentrated outside the network regions and appear to spread outwards and away from them, whereas the red excursions (including the downflowing RREs) in panel~(c) are mostly located on or within the boundaries of the strong network regions. 

\subsection{Statistical properties}
\label{Subsection:results-dimensions}
\begin{figure*}
   \centering
   \includegraphics[width=\textwidth]{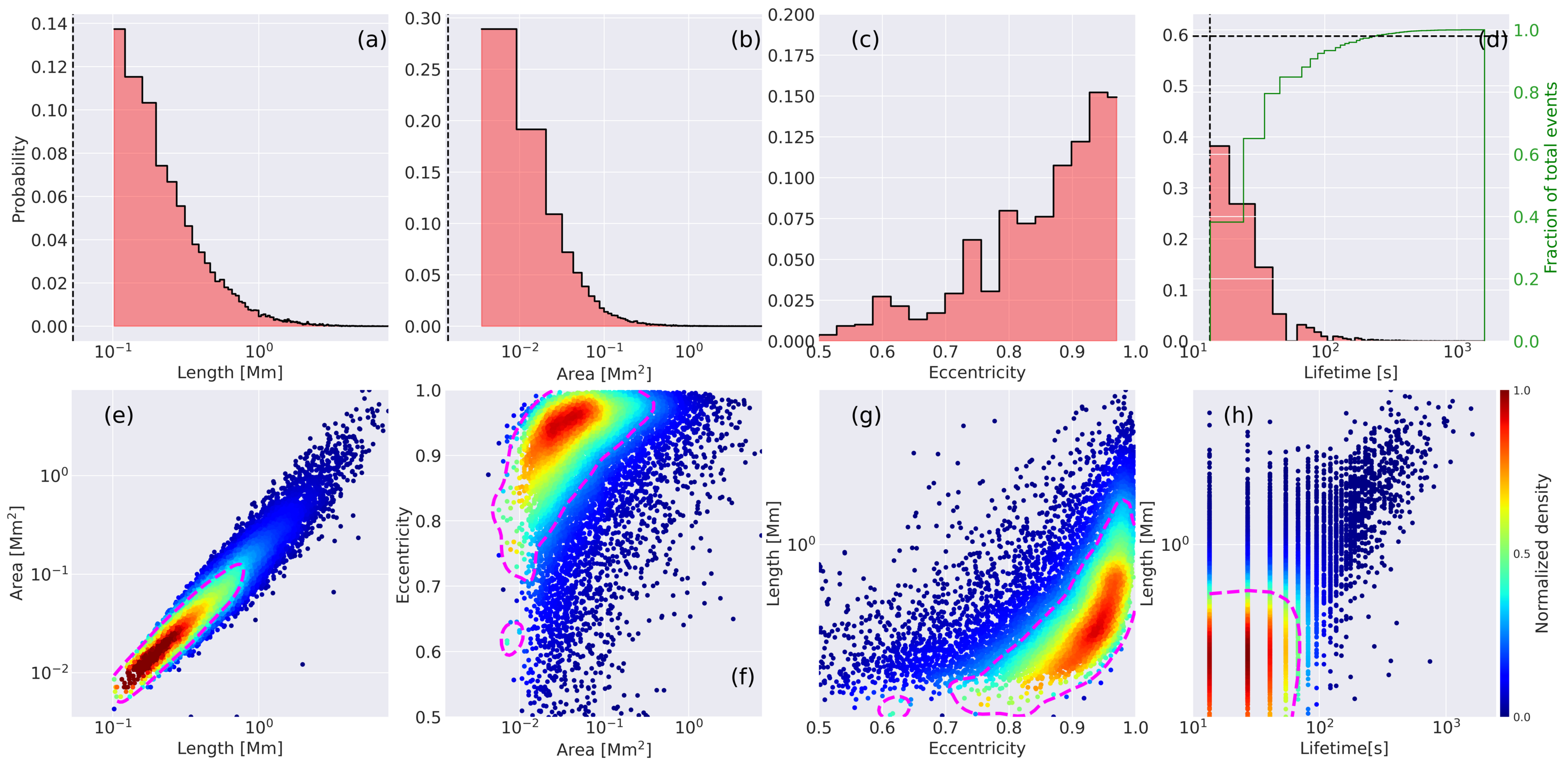}
   \caption{Dimensional analysis and lifetime statistics of RREs/downflowing RREs in our dataset. 1D histograms of length, area covered, eccentricity, and lifetime are shown in (a), (b), (c) and (d) respectively. Moreover, panel~(d) also shows the ECDF of the lifetime distribution in solid green and the dashed black horizontal line indicates the 98\% mark. The vertical dashed lines in panels (a), (b) and (d) indicates the spatial and temporal resolution limits of our data. Panels~(e)--(h) show the multivariate JPDFs between various quantities as indicated in a rainbow colormap with red (blue) indicating highest (lowest) density regions. The magenta contour overlaid on each of the density distribution indicates the region within which 70\% of the events lie.}
    \label{figure :Length_lifetime_stats_RREs}%
    \end{figure*}

\begin{figure*}
   \centering
   \includegraphics[width=\textwidth]{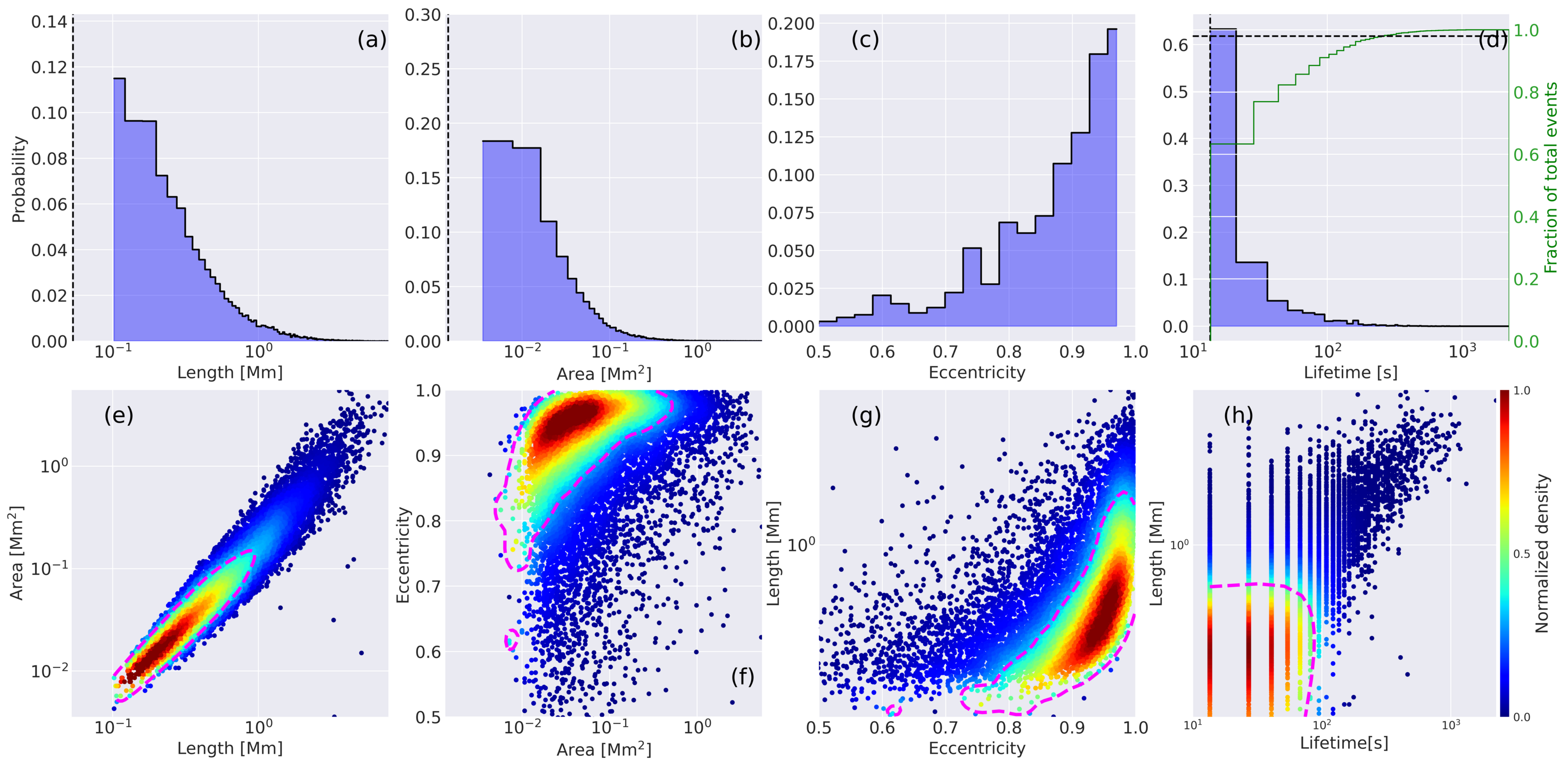}
   \caption{Same format as Fig.~\ref{figure :Length_lifetime_stats_RREs} above but for RBEs.
   }
    \label{figure:Length_lifetime_stats_RBEs}%
    \end{figure*}

Following the method discussed in Sect.~\ref{Subsection:method-stats}, we computed the maximum lengths, maximum areas, the corresponding eccentricities of the fitted ellipses and the lifetimes of 14,650 downflowing RREs/RREs and 19,643 RBEs. This section is dedicated to the description and comparison of the statistical properties of the newly reported downflowing RREs with respect to the traditional RBEs and RREs.

From the 1D histograms of length and area shown in panels~(a) and (b) respectively, of Figs.~\ref{figure :Length_lifetime_stats_RREs} and \ref{figure:Length_lifetime_stats_RBEs}, we find that the maximum length of the RBEs varies between 0.102~Mm to 7.8~Mm whereas for their red shifted counterparts, an upper limit of 7.75~Mm was found. In both cases a lower threshold length of 4 CHROMIS pixels was imposed which roughly translates to 0.1~Mm.
The area occupied by RBEs range from 0.003~Mm$^{2}$ to 5.53~Mm$^{2}$, while the RREs/downflowing RREs occupied an area that have a maximum value of 7.38~Mm$^{2}$, with roughly the same minimum. Both the area and length distributions appear to be skewed with a large number of data points clustered around 1~Mm.

Panel~(c) shows the 1D histogram of eccentricities of the fitted ellipses to the red and blue shifted excursions. A first glance at them shows that the events detected in both cases are rather elongated with $e$ $\ge$~0.5. This is well aligned to the known morphology of on-disk spicules that has been described in many studies in the past. Moreover, we also find that highly eccentric events tend to be more abundant. A close look at the eccentricity histograms, however, reveal that the distribution appears to be flatter for the RREs/downflowing RREs in Fig.~\ref{figure :Length_lifetime_stats_RREs}, compared to the RBEs in Fig.~\ref{figure:Length_lifetime_stats_RBEs}. This is further supported by the fact that the median value for the eccentricities corresponds to 0.85 for the former whereas for the latter it equates to 0.92. This indicates that, on average the red shifted excursions are slightly less elongated than their blue shifted counterparts. 

Panels~(e)--(g) shows the combined joint probability distribution functions (JPDFs) between various quantities as indicated. The magenta contours specify the regions within which 70~\% of the events lie. The motivation behind these JPDFs was to highlight the bi-variate relationships between the extracted quantities in the form of a 2D probability density function. The JPDF between maximum length vs. maximum area of the blue shifted and red shifted events in panel~(e) show a strong correlation indicating that spicules that occupy larger area are also longer. Further investigation from panels~(f) and (g) reveal that the bulk (in this case 70~\% or more) of the events are highly eccentric and are less than 1~Mm in length and 0.5~Mm$^{2}$ in area. Also in general, events that are more eccentric also tend to be larger in size. This is true for both the RBEs and RREs/downflowing RREs. 

Panels~(d) and~(h) in Figs.~\ref{figure :Length_lifetime_stats_RREs} and \ref{figure:Length_lifetime_stats_RBEs} show the 1D distributions and the JPDFs between lifetime and length of downflowing RREs/RREs and RBEs, respectively. We find RBEs lasting from 13.6~s to 2140~s, and RREs/downflowing RREs lasting from 13.6~s to 1618~s with a median lifetime of about 27.2~s in both cases. Moreover, we also show an empirical cumulative distribution function (ECDF) of the lifetime that clearly indicates that 98\% of the events are shorter than 200~s. 
%
%
The events lasting longer than $\ge$~300~s in the above distributions make up less than 2\% in our data and can be considered as outliers. Part of these are related to a few surge-like events that are morphologically quite different from type-II spicules but can have similar spectral properties. It can further not be excluded that some of these are actually multiple recurring spicules that the method cannot separate and classifies as single long duration events.
The JPDFs in panel~(h) also show that more than 70\% of the events that last shorter than 100~s, are also shorter than 0.7~Mm. The results described above complies well with the characteristic properties of type-II spicules and are well in agreement with the values reported in earlier works such as \citet{Luc_2009,Sekse_2012} and \citet{Tiago_2012}. Further analyses showing the relationship between the eccentricity of spicules vs. their lifetimes are shown in Fig.~\ref{figure:appendix-combined_eccentricity} in appendix~\ref{appendix-supp-figs}, which reinforces that the majority of the detected events are both highly eccentric and rapid.

The results presented above do not explicitly distinguish between the downflowing and traditional RREs, but they strongly suggest that the newly reported downflowing RREs have properties similar to their upflowing counterparts observed in both the blue ward and red ward side of the \halpha\ line core. This further reinforces our proposition that the downflowing RREs belong to the family of type-II spicules with an opposite apparent motion.

\section{Discussion}
\label{Section:Discussion}
\subsection{Interpretation of the downflowing RREs}
\label{subsection:interpretation_DRRE}

The examples shown in Fig.~\ref{figure:DRREs} and the statistical analysis of the proper motion of stronger and weaker excursions described in Fig.~\ref{figure:zoom-in-apparent-motion}, indicate that there exists a new category of RREs that behave contrary to the traditional interpretation of RREs. \citet{Bart_3_motions,2013ApJ...769...44S,2014Sci...346D.315D} and \citet{2015ApJ...802...26K} explained that RREs, like RBEs, are a manifestation of the same physical phenomenon and appear either in the blue or red wing of \halpha\ depending on their transverse motion along the LOS of chromospheric magnetic field lines. In other words, RREs are generally found in close association with RBEs. The downflowing RREs, discussed in this paper, do not satisfy this explanation since they are not often found associated with their blue wing counterparts. Moreover, their opposite apparent direction of motion indicates that they do not occur due to the swaying and torsional motions often associated with type-II spicules \citep{Bart_3_motions}. 

The motion associated with chromospheric spicules is quite complex and often all the three, i.e., the field aligned flows, the torsional motions and the transverse swaying motions, are at play. So far, the appearance of isolated RBEs are interpreted to be the result of upflows along the field lines parallel to the spicule axis and the RREs the result of the combination of the three. The appearance, morphology and their statistical properties discussed in this paper suggests that downflowing RREs are similar to RBEs and they are the result of the field aligned \textit{downflows} in the solar chromosphere. 

These downflows, however, are quite different from the type-I spicules such as active region dynamic fibrils or quiet sun mottles that are commonly observed near the \halpha{} line core images \citep{Luc_2007}. The $\lambda$t slices shown at the bottom of Fig.~\ref{figure:DRREs} show that the downflowing RREs have no a priori blue shift associated with them, a characteristic that is often linked with type-I spicules. Moreover, various studies such as \citep{Bart_2007_PASJ,Tiago_2012} show that type-Is have much lesser LOS velocities (Doppler offsets) and seldom appear at wing positions so far from the line core. Furthermore, the morphological appearance of the downflowing RREs are very similar to RBEs and RREs, which also provides a strong evidence that the former category is not associated with the red shifts of the fibrils but are more like the downflowing counterparts of RREs.

\subsection{What might cause downflowing RREs?}
\label{Subsection:cause_DRREs}

One of the most important questions that remains to be addressed is what might be the origin of these downflowing RREs? In this section we discuss some of the possible mechanisms that could be responsible for such ubiquitous downflows.

Type-II spicules sometimes exhibit parabolic up-down motion much like the dynamic fibrils \citep{Tiago_2014_heat}. However, unlike the latter, the type-II spicules are much faster, shorter lived (in the chromosphere) and show signatures in the transition region (TR) and even in the solar corona \citep{2011Sci...331...55D,Vasco_2016,2019Sci...366..890S}. During their ascending phase they get rapidly heated to TR with signatures in passbands sampling coronal (1~MK) temperatures, in both active regions and the quiet Sun, and eventually fall back in the chromosphere. These downflowing plasma could, in principle, be responsible for the observed downflowing RREs in the \halpha{} red wing. The combined upflowing and downflowing lifetimes in such cases can well be above 600--700~s \citep{Tiago_2014_heat,2019Sci...366..890S} which makes them typically last longer than their purely chromospheric type-I counterparts (lifetime \textasciitilde 3--5 min).

A supporting proposition in favor of this returning plasma in the spicular form can stem from the fact that observations of the spectral lines formed in the temperature regimes between \textasciitilde15,000~K and 2.5 $\times$ 10$^{5}$~K, reveal the prevalence of an average red shift or downflowing motion of the order of $10$--$15$~\kms{} in the TR \citep{1976ApJ...205L.177D,1981ApJ...251L.115G,1999ApJ...522.1148P,2011A&A...534A..90D} as discussed in Sect.~\ref{Section:intro}. These studies imply the existence of plasma flows or wave motions in the quiet Sun with amplitudes that are significant fractions of the sound speed in the TR.

Different mechanisms have been proposed in the past to explain the net downflowing structure of the TR. \citet{2006ApJ...638.1086P} synthesized spectra of the coronal and TR lines with the help of a 3D numerical simulation spanning the photosphere to the corona and showed that the persistent red shifts in the TR could possibly be explained by the effect of the flows caused due to heating by magnetic braiding. \citet{2010ApJ...718.1070H} expanded on this model and injected an emerging flux to the 3D model of \citet{2006ApJ...638.1086P} and concluded that these downflows are a result of the rapid episodical heating between the upper chromosphere and lower corona. \citet{2018A&A...614A.110Z} further complemented these earlier studies by establishing that pressure driven downflows along the magnetic field lines could be identified as one of the key mechanisms responsible for these observed red shifts in the TR. 

Perhaps the most likely mechanism applicable to the observations of downflowing RREs is the proposition that TR red shifts are caused due to the emission from the return flows, that had been formerly heated and injected into the solar corona by spicules \citep{1977A&A....55..305P,1982ApJ...255..743A,1984ApJ...287..412A}. Based on their modeling efforts, the studies by the above sets of authors estimated that spicular material heated to coronal temperatures, carry a large upward flux (upflows) that is almost 100 times the flux measured due to solar wind at 1 AU. Consequently, they reasoned that the rest of the mass must return to solar atmosphere and it should happen in the form of spicular downflows which are then in turn responsible for the observed net red shifts in the TR. Older numerical calculations such as the one by \citet{1987ApJ...319..465M}, however, could not obtain these observed red shifts. Moreover, many earlier numerical models could not even account for any blue shift (upflows) in the coronal or TR passbands. Therefore, the community at large remained skeptical about the contribution of spicules in these downflows. Recent observations by \citet{2009ApJ...707..524M,Luc_2015} and  state-of-the-art numerical modeling efforts by \citet{Juan_2017_Science,Juan_2018} addressed these concerns and proved beyond doubt that spicules harbor a clear blue shift when observed in hotter TR and coronal spectral channels.

The observations of these downflowing features have mainly been limited to the TR till date. None of the studies in the past found evidences of the chromospheric counterparts of these TR downflows. In some studies \citep[such as][]{1987ApJ...319..465M} the non existence of the chromospheric signatures of the TR downflows raised serious doubts on their impact in spicular plasma in the past. However, the examples of the ubiquitous downflows and the detailed analysis presented in this paper seem to suggest that the downflowing RREs could well be the signatures of the TR downflows in the chromosphere. Whether they follow a typical parabolic trajectory or not warrants further investigation but our results strongly revives the possibility that returning spicular materials can be one of major driving mechanisms for the observed TR red shifts and downflowing RREs.

Alternate possible explanation for the origin of the downflowing RREs could lie in the investigation carried out by \citet{2019A&A...632A..96R} where they conjectured that RBEs could display return flows in the form of cool plasma that traces the trails of the preceding type-II spicules. Their statistical analysis strongly suggests that there is a tight correlation between the occurrences of RBEs and subsequent \halpha{} fibrils within a certain time delay. Consequently, they attribute the dark fibrilar appearance around the chromospheric network regions (seen predominantly in \halpha{}) mainly to preceding type-II spicules. The downflowing RREs, like the ones presented in this paper, could well be the immediate following state of these firbils as they continue to evolve. However, so far there are no studies that could firmly establish this relation.

Finally, the chromospheric counterparts of the low lying loops found in the TR 
\citep{2014Sci...346E.315H, 
2018A&A...611L...6P} could also be attributed to the appearance of these downflowing RREs in \halpha{}. The low lying loops are not found to show prominent signals in the \halpha{} line wing images, except in the far wings, including the red wing. If such red-wing signatures are true flows then a footpoint could appear as a downflowing RRE whereas most of the loop would be "hiding" in TR passbands.  The footpoints of nearly all the loops observed by 
\citet{2018A&A...611L...6P} 
lie in close proximity or share the footpoints of chromospheric spicules predominantly rooted in the network regions which makes them possible to have a relation with the downflowing RREs as described above. Furthermore, \citet{2020ApJ...889...95M} also discussed the possibility of spicules forming along the loops in numerical simulations which could be further compared with the observations. However, given the fact that these loops are not as ubiquitous as spicules on the solar surface it is very unlikely that downflowing RREs are always associated with the chromospheric signatures of these low lying TR loops.

\subsection{Comparison with flocculent flows}
\label{Subsection:Flocculence}

Flocculent flows in the solar chromosphere were first described by \citet{2012ApJ...750...22V} as distinct small-scale features that move
intermittently towards and away from a sunspot. They bear morphological resemblance to coronal rain, both qualitatively and quantitatively \citep{2012ApJ...745..152A}, but their sizes are somewhat smaller and they move at much lower average velocities and over shorter distances.
Flocculent flows have been suggested to be a result of a siphon flow driven by pressure difference between the footpoints in a loop.  
Compared to downflowing RREs, flocculent flows typically travel over larger distances and consist more of distinct and isolated blob-like features. 
We observe downflowing RREs exclusively in close vicinity to magnetic network areas while flocculent flows were observed to travel along extended parts of the sunspot superpenumbra and long active region fibrils.



\subsection{Spatial distribution of spicules and their substructures}
\label{subsection:spatial_dist_discussion}

The results presented in Fig.~\ref{figure:unique_spicules_velocity} suggests that, on average, the stronger excursions in both the redward and blueward side of \halpha\ line core are located closer to the strong magnetic field regions, whereas the weaker counterparts appear to be scattered across the FOV. This seems to be the case for both the red and blue shifted excursions. Moreover, the weaker excursions occupy lesser area on the FOV on an individual event-by-event basis. Since most of the events in the stronger red excursion category have apparent inward motion (refer to Figs.~\ref{figure:zoom-in-apparent-motion} and \ref{figure:appendix-full-fov-apparent-motion}), we conclude that on average the downflowing RREs are predominantly located in the regions close to the network areas, whereas their traditional counterparts appear to exist slightly farther away. 

The spicule halos, described in Sect.~\ref{Subsection:shadows}, are a consequence of the substructures seen in RBEs and RREs/downflowing RREs. These substructures are mainly due to the fact that spicules often have a large distribution of spectral properties such as line core width, Doppler shift \citep{2016ApJ...824...65P} or absorption in the wings of the \halpha{} spectral line, not only along the spicule axis but also in the transverse direction, and are a part of the same morphological structure that evolves in a collective fashion. Figure~\ref{figure:shadows} and its associated animation strongly suggests that many times spicules do not evolve independently but in groups that maybe difficult to discern when observed at one wavelength position. This reinforces the fact that group behavior is common among type-II spicules \citep{2014ApJ...795L..23S}. 

Another important feature commonly observed in spicules is the rapid morphological transformations that they undergo both in space and time, including multiple splitting and branching \citep{2020ApJ...891L..21Y}. We refer to the examples shown in Fig.~\ref{figure:DRREs} where we clearly see that the DRRE-1 initially starts as one structure but towards the end of its evolution we see it clearly split into two as they terminate at the network bright points. RBE-1 also shows clear signs of morphological branching with a "Y" pattern just before it disappears. Often these branches have different LOS velocities but as we see in the examples they are clearly associated with the original parent structure evolving together. Therefore it implies that, downflowing RREs, like RBEs and RREs, can have multiple structures either in situ, due to a range of spectral properties, or due to the morphological transformations they undergo during their evolution.

\subsection{Significance of the detection technique and their limitations}
\label{subsection:stats_properties_discussions}

The detection method based on $k$-means clustering and morphological image processing technique presented in this paper enabled us to study and analyze spicules in an unprecedented detail. Identifying the RBE and RRE/downflowing RRE RPs on the basis of the strength of their Doppler offset, line core width and enhanced absorption measure in the \halpha\ spectral line provides one of the most compendious approach in their characterization. 
 
Nearly 20,000 RBEs and 15,000 RREs (including downflowing RREs) have been identified in our 97~minute long dataset that allowed us to perform varied statistical analysis and compare the properties of the downflowing RREs in the context of traditional RBEs and RREs. Though no distinction was possible between the downflowing RREs and RREs in terms of their spectral signatures, we see that out of the 15,000 red shifted events a substantial number of strongest excursions are downflowing in nature, with their apparent motion directed towards the strong network areas. Furthermore, the statistical analysis presented in this paper reveals that downflowing RREs have similar dimensions and lifetimes when compared against RBEs and traditional RREs. This makes them more likely to be a part of the family of type-II spicules.

Previous reports \citep{Luc_2009,Sekse_2012,Tiago_2012,2015ApJ...802...26K} found comparatively longer lengths for RBEs and RREs. This could possibly be due to the fact that, except for \citet{Tiago_2012}, most of these earlier works have focused on performing the statistical analyses based on images at single wavelength positions far in the blue or the red wing of the \halpha{} line profile. Moreover, on most occasions they employed a lower length detection threshold of \textasciitilde $0.730$~Mm that strictly limited the lower estimate of their analysis, thereby facilitating the detections of relatively longer events. On the other hand, the method employed in this paper is compendious both in terms of spectra as well as in size thereby enabling smaller detections.

The technique presented in this paper exploits the complete spectral profile (RPs) instead of relying on the morphology of spicular features at single wavelength positions. To get rid of erroneous detections we also impose a lower limit on the length of the detected events but it is far lower (\textasciitilde 0.1~Mm) than the ones chosen in the earlier works. \citet{Luc_2009} reported the presence of special RBEs in the form of "black beads" in \halpha{} that appeared as tiny roundish darkenings and were interpreted as spicules that were aligned closely along the LOS. The dimension of such features were reportedly between 0.15 and 0.3~Mm. Interestingly, \cite{2010PASJ...62..871A} reported the shortest mean lengths of spicules compared to all other works in the recent past. \cite{Tiago_2012} interpreted this discrepancy owing it to the exponential drop in intensity along the body of the spicule which can possibly render the tops too faint to be picked up in their detection when viewed against the disk. This explanation is also valid in our case because, despite the spatial resolution, the top portion of the spicules most likely has too little opacity to be considered as a part of either an RBE or an RRE/downflowing RRE. Therefore, there could be an intrinsic bias in determination of the lengths of spicules as was also reported by \citet{Sekse_2012}. The lifetimes, on the other hand, are well in agreement with most of the former studies mentioned above, despite the increased capture of events over the FOV and increased sensitivity to smaller scales. 

Despite undertaking advanced methods to characterize and detect RBEs and RREs, we do encounter a few limitations. In this paper, we have referred to the detected events as RBEs or RREs/downflowing RREs which are on-disk counterparts of type-II spicules; but are we clearly detecting the type-IIs? One of the limitations of $k$-means clustering is that it only accounts for the spectral lines in our case. Therefore, both type-I and type-II spicules could very well be included in our detections as they have similar spectral profiles in \halpha{}. However, the results from the detailed statistical analysis presented in this paper strongly suggest that we are in the domain of type-II spicules since, most importantly, 98\% of the events last under \textasciitilde$3$~minutes and the median lifetime is about $27$~s. Furthermore the morphology and the evolution of the detected events, like the ones indicated in Figs.~\ref{figure:sample_detection} and \ref{figure:shadows}, clearly suggest that we are detecting the rapid type-II spicules.


\section{Concluding remarks}
\label{Conclusion}

We report the first observation and characterization of rapid downflows in the solar chromosphere in the form of spicules. Their rapidity and apparent motion in the far red wing images of \halpha{} strongly suggests that they are the downflowing counterparts of the traditional RREs first reported by \citet{2013ApJ...769...44S}, unlike coronal rain \citep{2012ApJ...745..152A} and flocculent flows \citep{2012ApJ...750...22V}. We therefore term them downflowing RREs. In depth statistical analyses performed on 14,650 RREs and 19,643 RBEs imply that downflowing RREs are similar to the already known RREs and RBEs. Moreover, they also undergo rapid morphological transformations during their evolution in the same way as RBEs and RREs. The only evident difference of this new class of RREs lies in their apparent motion where they seem to originate away and terminate close to the strong field network regions. This suggests that they are a result of the field aligned downflows in the solar chromosphere. Furthermore, the downflowing RREs could also undergo the transverse and  torsional motions that are often associated with type-II spicules which hints at the possibility of finding \textit{downflowing RBEs} in the solar chromosphere. Future work in this direction could shed more light in this context.

The downflowing RREs could possibly be linked to the return flows of type-II spicules which ascend rapidly from the chromosphere to the TR and even coronal heights, and eventually fall back. Moreover, we present arguments suggesting that these downflowing RREs could well be responsible for the observed TR downflows as was first hypothesized by \citet{1977A&A....55..305P}. It would be interesting to find direct signatures of these downflowing RREs in the corona and the TR with coordinated observations from the \textit{Interface Region Imaging Spectrograph} and \textit{Solar Dynamics Observatory} because it will enable better understanding of the mass and energy cycle in the solar atmosphere. Such a study is currently underway and will be the subject of a forthcoming paper.

\begin{acknowledgements}
We thank Ainar Drews for his help with the observations. The Swedish 1-m Solar Telescope is operated on the island of La Palma by the Institute for Solar Physics of Stockholm University in the Spanish Observatorio del Roque de los Muchachos of the Instituto de Astrof\'{i}sica de Canarias. The Institute for Solar Physics is supported by a grant for research infrastructures of national importance from the Swedish Research Council (registration number 2017-00625). This research is supported by the Research Council of Norway, project number 250810, and through its Centers of Excellence scheme, project number 262622. VMJH is also funded by the European Research Council (ERC) under the European Union’s Horizon 2020 research and innovation programme (SolarALMA, grant agreement No. 682462).
\end{acknowledgements}

\bibliographystyle{aa} 
\bibliography{spicules_4.bib} 

\begin{appendix}

\section{Spicule substructures}
\label{Appendix:substructures}
\begin{figure*}[htb!]
  \centering
  \includegraphics[width=\textwidth]{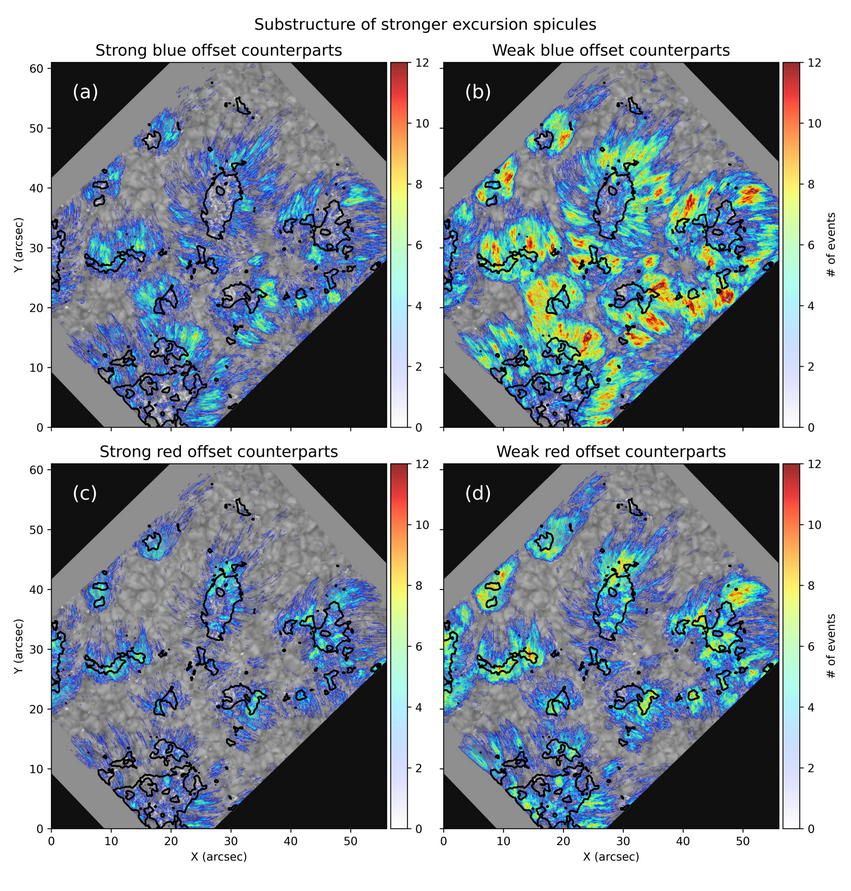}
  \caption{Discerning the substructures of the RBEs, RREs and, downflowing RREs but only for the spicules harboring at least one of strong excursion RPs (2, 3, 6 and 7). Panels (a) and (b) show the strong (RPs~2 and 3) and weak (RPs~0 and 1) offset counterparts of the stronger excursion RBEs, respectively; whereas panels (c) and (d) shows the strong (RPs~6 and 7) and weak (RPs~4 and 5) offset counterparts belonging to the stronger RREs/downflowing RREs. The color gradient indicates the density of the events detected over the entire duration of the data. The black contour indicates the regions that have an absolute value of the LOS magnetic field $\geq$ $100$~G.
  }
    \label{figure:appendix-same_spicules_different_velo}%
    \end{figure*}

We describe the substructures of spicules belonging to the stronger excursions in blueward and redward side of \halpha{} line core. We reiterate that both the stronger and weaker excursion spicules can have spatial substructures due to variation in their spectral properties which causes them to have different RPs (stronger excursion RPs along with weaker excursion RPs).
As described in Sect ~\ref{subsubsection:Identifying-RPs} and ~\ref{subsection:DRREs-observation}, we considered those spicules as stronger excursion that have at least one of stronger excursion RPs, i.e., RPs 2, 3, 6, and, 7, at least once during their entire lifetime. 
However, as discussed earlier in the paper, even the strongest excursions have a spatial variation in their Doppler offset along and perpendicular to the spicule axis, hence weaker RPs, i.e., RPs 0, 1, 4 and, 5, could also be present within their morphological structure (see Sect.~\ref{Subsection:shadows}).
The major motivation of the analysis presented in this section is to discern the substructures of strong RBEs, RREs, and downflowing RREs depending on RPs and see if stronger and weaker excursion RPs have any spatial preference.     
These variations add to the structural complexity of spicules as shown in Fig.~\ref{figure:appendix-same_spicules_different_velo}. Panel~(b) shows the "weaker" Doppler offset counterparts of the stronger blue excursion events in panel~(a). The arrangement is identical for the red excursions in panels~(c) and (d). Therefore, unlike Sect.~\ref{Subsection:spatial-distribution-spicules}, only the events common to the stronger excursion category are identified and represented in panels~(b) and (d) of Fig.~\ref{figure:appendix-same_spicules_different_velo}, while all others are excluded. Consequently, Fig.~\ref{figure:unique_spicules_velocity} shows the distribution of spicules belonging uniquely to the strong and weak excursion categories whereas Fig.~\ref{figure:appendix-same_spicules_different_velo} shows the spatial distribution of the spicules belonging only to the stronger excursion category, but with different excursion RPs. Naturally, the events in the different velocity compartments are now no longer unique.   

Upon closer investigation, we find that the number density of events increases roughly by 45\% for the strong blue excursions (panels (a)--(b)) and 27\% (panels~(c)--(d)) for the strong red excursions. Furthermore, this increase is predominantly reflected in the regions away from the strong network areas in the former, whereas for the red excursions this is also linked to an increase within and inside the strong field contours such as within the network patch located at ($X$,$Y$) =(45\arcsec,33\arcsec) or close to the bottom of the FOV at $X$=20\arcsec in panel~(d). The increase in the number density can be attributed to the variation in the Doppler offsets, and line widths of spicules in different parts of their body.

\section{Supplementary figures}
\label{appendix-supp-figs}
This section provides supplementary resources for the results presented in Sects.~\ref{Subsection:method-morphology}, \ref{subsection:DRREs-observation} and \ref{Subsection:results-dimensions}. The details of the figures shown in this section are discussed in the main text. 

\begin{figure*}[htb!]
  \centering
  \includegraphics[width=\textwidth]{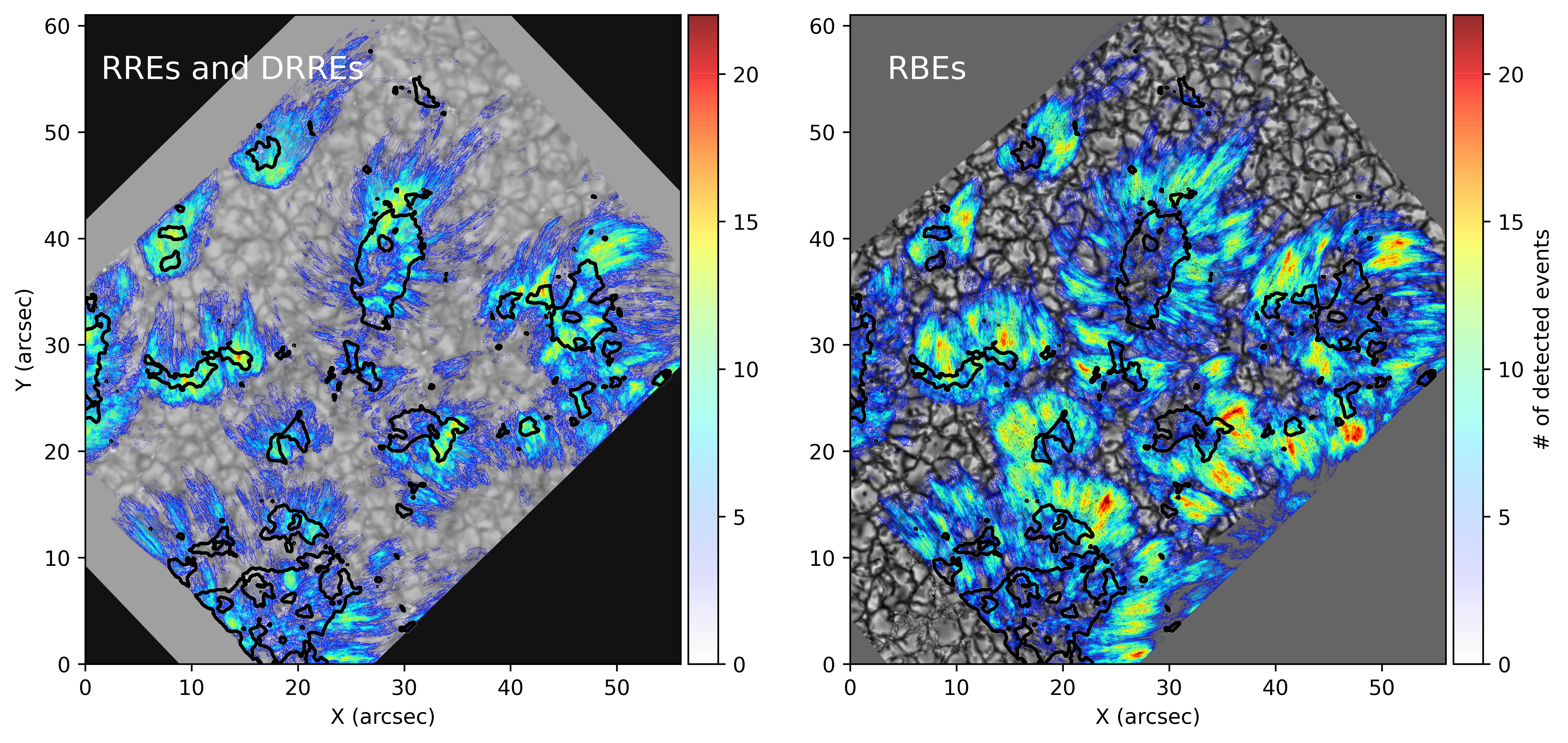}
  \caption{Overview of the location and density distribution of the total detected events over the whole FOV. Panel~(a) shows the occurrence of RREs/downflowing RREs over the whole time series against an \halpha{} red-wing image at 93~\kms; and panel~(b) shows the occurrence of RBEs against a background of CHROMIS continuum image at 4000~\AA.}
  \label{figure:appendix-spicules}%
\end{figure*}

\begin{figure*}[htb!]
  \centering
  \includegraphics[width=0.76\textwidth,keepaspectratio]{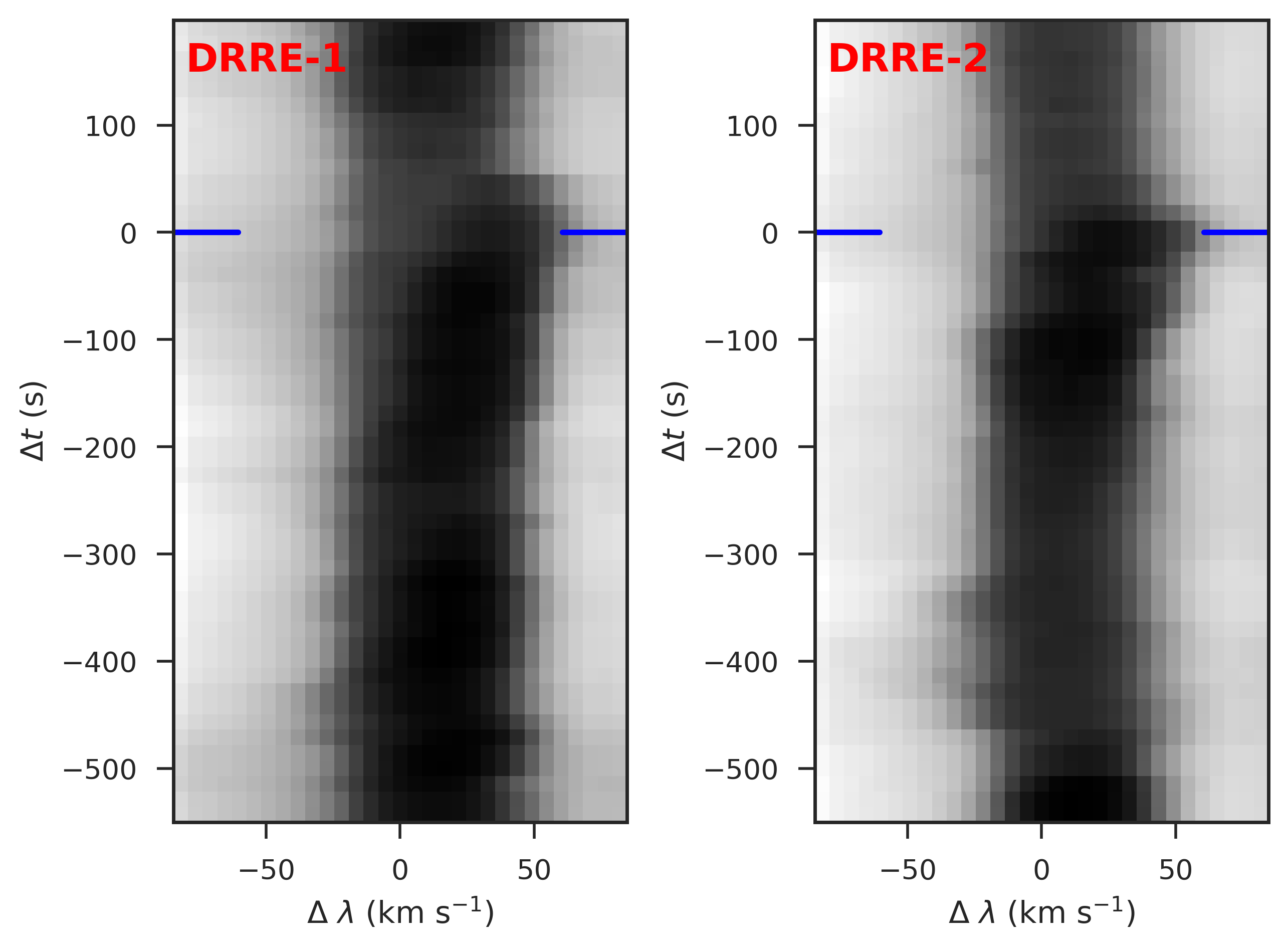}
  \caption{Extended $\lambda t$ diagrams corresponding to the two downflowing RREs shown in Fig.~\ref{figure:DRREs} displaying the lack of any blue shift preceding the downflowing RREs. The maximum excursion towards the red-wing of \halpha{} is indicated by blue horizontal markers which also corresponds to the time shown in the $\lambda t$ diagram of Fig.~\ref{figure:DRREs}.}
  \label{figure:appendix-lambda-t-extended}%
\end{figure*}

\begin{figure*}[htb!]
  \centering
  \includegraphics[width=\textwidth]{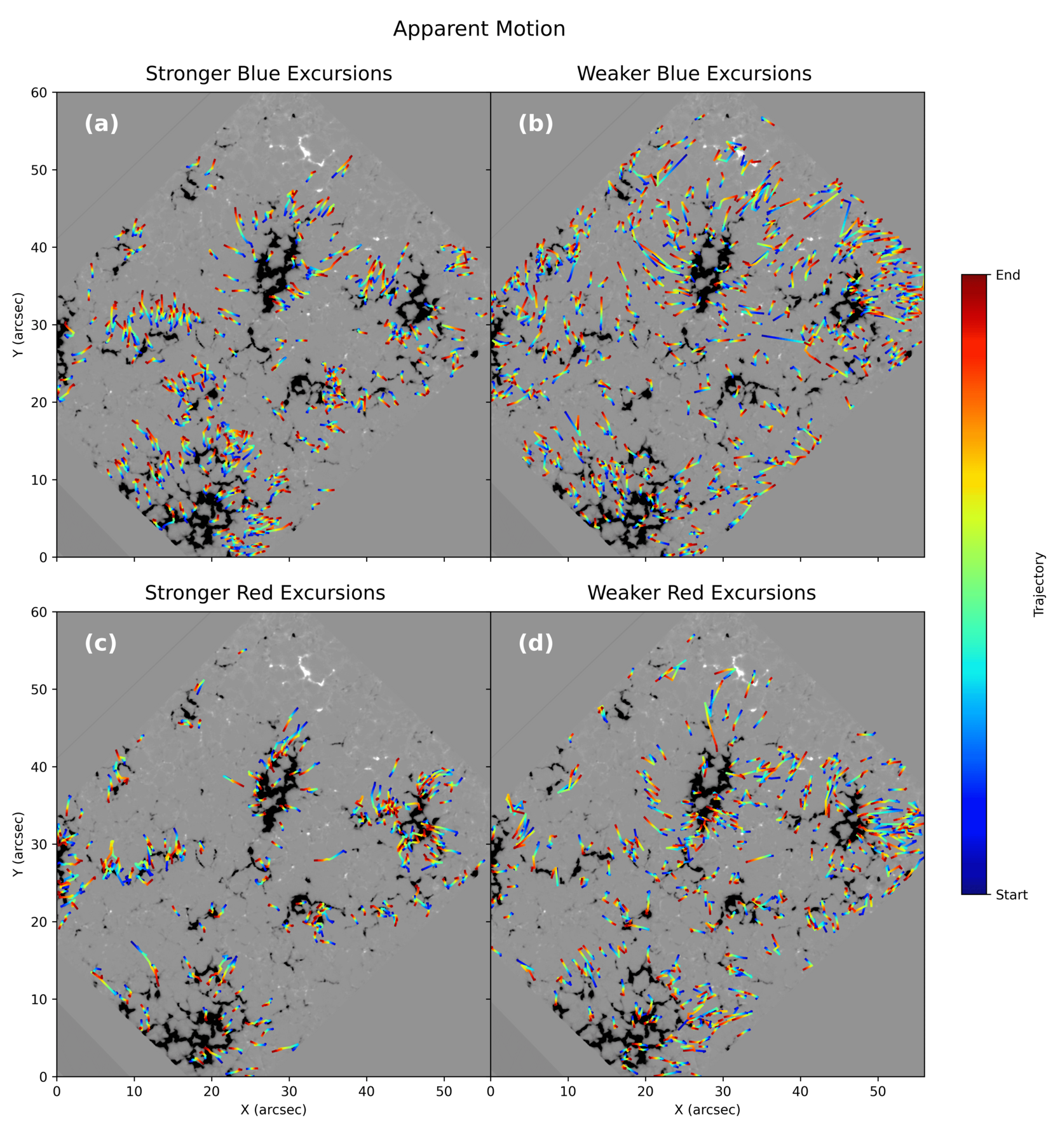}
  \caption{Apparent motion of spicules in the same format as Fig.~\ref{figure:zoom-in-apparent-motion} but for the full FOV. The behavior explained in the text is also seen to be followed all over the FOV. Displacements smaller than 1\arcsec have been removed in order to improve the visibility.
  }
  \label{figure:appendix-full-fov-apparent-motion}%
\end{figure*}

\begin{figure*}[htb!]
  \centering
  \includegraphics[width=\textwidth]{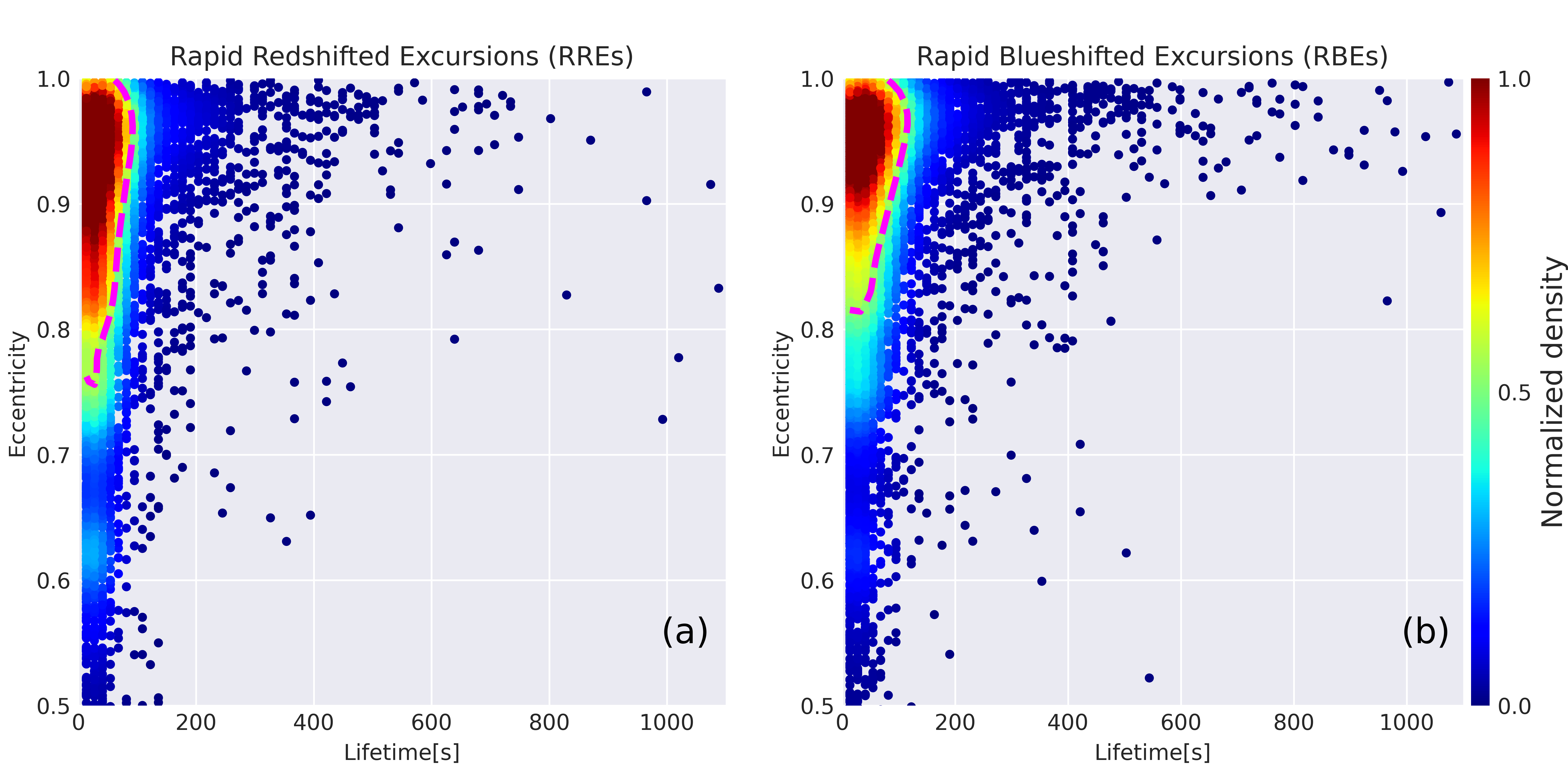}
  \caption{Bi-variate joint probability density between lifetime and eccentricity (of the fitted ellipses) of (a) RREs/downflowing RREs and (b) RBEs in rainbow colormap as in Fig.~\ref{figure:Length_lifetime_stats_RBEs}. The magenta contour indicates the region within which 70~\% of the events lie.  }
  \label{figure:appendix-combined_eccentricity}%
\end{figure*}

\end{appendix}
\end{document}